\documentclass[]{pasj01}

\usepackage[Symbolsmallscale]{upgreek}

\newcommand{\eg}{e.g., }
\newcommand{\ie}{i.e., }
\newcommand{\Msun}{M_{\odot}}

\newcommand{\RADEC}{${\rm R.A.=09^{h}40^m50^s}$, ${\rm decl.=-03^\circ27'05''}$ (J2000.0)}
\newcommand{\FRB}{FRB~151230}
\newcommand{\Ebvh}{E_{B-V, {\rm host}}}
\newcommand{\Ebvg}{E_{B-V, {\rm Gal}}}
\def\gsim{\mathrel{\rlap{\lower 4pt \hbox{\hskip 1pt $\sim$}}\raise 1pt
\hbox {$>$}}}
\def\lsim{\mathrel{\rlap{\lower 4pt \hbox{\hskip 1pt $\sim$}}\raise 1pt
\hbox {$<$}}}

\newcommand{\SNa}{Cand-2}
\newcommand{\SNb}{Cand-3}
\newcommand{\SNc}{Cand-6}
\newcommand{\SNd}{Cand-11}

\newcommand{\SNg}{Cand-5}
\newcommand{\texp}{t_{\rm exp}}
\newcommand{\MB}{M_B}
\newcommand{\nobs}{n_{\rm obs}}

\begin{document} 
\Received{2018/04/19}
\Accepted{2018/08/09}

\title{Optical follow-up observation of Fast Radio Burst 151230}

\author{Nozomu \textsc{Tominaga}\altaffilmark{1,2}}
\author{Yuu \textsc{Niino}\altaffilmark{3}}
\author{Tomonori \textsc{Totani}\altaffilmark{4,5}}
\author{Naoki \textsc{Yasuda}\altaffilmark{2}}
\author{Hisanori \textsc{Furusawa}\altaffilmark{3}}
\author{Masayuki \textsc{Tanaka}\altaffilmark{3}}
\author{Shivani \textsc{Bhandari}\altaffilmark{6,7,8}}
\author{Richard \textsc{Dodson}\altaffilmark{9}}
\author{Evan \textsc{Keane}\altaffilmark{10}}
\author{Tomoki \textsc{Morokuma}\altaffilmark{11}}
\author{Emily \textsc{Petroff}\altaffilmark{12}}
\author{Andrea \textsc{Possenti}\altaffilmark{13}}
\altaffiltext{1}{Department of Physics, Faculty of Science and Engineering, Konan University, 8-9-1 Okamoto, Kobe, Hyogo 658-8501, Japan}
\altaffiltext{2}{Kavli Institute for the Physics and Mathematics of the Universe (WPI), The University of Tokyo Institutes for Advanced Study, The University of Tokyo, 5-1-5 Kashiwa, Chiba 277-8583, Japan}
\altaffiltext{3}{National Astronomical Observatory of Japan, 2-21-1 Osawa, Mitaka, Tokyo 181-8588, Japan}
\altaffiltext{4}{Department of Astronomy, School of Science, The University of Tokyo, 7-3-1 Hongo, Bunkyo-ku, Tokyo 113-0033, Japan}
\altaffiltext{5}{Research Center for the Early Universe, School of Science, The University of Tokyo, 7-3-1 Hongo, Bunkyo-ku, Tokyo 113-0033, Japan}
\altaffiltext{6}{CSIRO Astronomy and Space Science, PO Box 76, Epping, NSW 1710, Australia}
\altaffiltext{7}{Centre for Astrophysics and Supercomputing, Swinburne University of Technology, Mail H30, PO Box 218, VIC 3122, Australia}
\altaffiltext{8}{ARC Centre of Excellence for All-sky Astrophysics (CAASTRO), Australia}
\altaffiltext{9}{International Centre for Radio Astronomy Research,  The University of Western Australia, 35 Stirling Hwy, Western Australia}
\altaffiltext{10}{SKA Organisation, Jodrell Bank Observatory, Macclesfield, Cheshire, SK11 9DL, UK}
\altaffiltext{11}{Institute of Astronomy, Graduate School of Science, The University of Tokyo, 2-21-1 Osawa, Mitaka, Tokyo 181-0015, Japan}
\altaffiltext{12}{ASTRON Netherlands Institute for Radio Astronomy, Oude Hoogeveensedijk 4, 7991PD Dwingeloo, The Netherlands}
\altaffiltext{13}{INAF-Osservatorio Astronomico di Cagliari, Via della Scienza 5, I-09047 Selargius (CA), Italy}
\email{tominaga@konan-u.ac.jp}



\maketitle

\begin{abstract}
 The origin of fast radio bursts (FRBs), bright millisecond radio
 transients, is still somewhat of a mystery. Several theoretical models
 expect that the FRB accompanies an optical afterglow (\eg \cite{tot13,kas13FRB}). In order to investigate
 the origin of FRBs, we perform $gri$-band follow-up
 observations of \FRB\ (estimated $z \lsim 0.8$) with Subaru/Hyper Suprime-Cam at $8$, $11$,
 and $14$~days after discovery. The follow-up
 observation reaches a $50\%$ completeness magnitude of $26.5$~mag for
 point sources,
 which is the deepest optical follow-up of
 FRBs to date. We find $13$ counterpart candidates
 with variabilities during the observation. We investigate their
 properties with multicolor and multi-wavelength observations and
 archival catalogs. Two candidates are excluded by the
 non-detection of \FRB\ in the other radio feed horns that operated
 simultaneously to the detection, as well as the inconsistency between the
 photometric redshift and that derived from the
 dispersion measure of \FRB. Eight further candidates are
 consistent with optical variability seen in AGNs. Two more
 candidates are well fitted with transient templates (Type IIn
 supernovae), and the final candidate is poorly fitted with all of our
 transient templates and is located
 off-center of an extended source. 
 It can only be reproduced with rapid transients with a faint peak and
 rapid decline and the probability of chance coincidence is 
 $\sim3.6\%$. We also find that none of our candidates are consistent with
 Type Ia supernovae, which rules out the association of Type Ia
 supernovae to \FRB\ at $z\leq0.6$ and limits 
 the dispersion measure of the host galaxy to 
 $\lsim300$~pc~cm$^{-3}$ in a Type Ia supernova scenario.

\end{abstract}

\section{Introduction}
\label{sec:intro}

Fast radio bursts (FRBs) are bright millisecond radio transients that
were first discovered in a search of archival data of the Parkes Radio
Telescope \citep{lor07}. Subsequent searches discovered further
FRBs with the Parkes Radio Telescope (\eg \cite{cha16}),
the Green Bank Telescope \citep{mas15}, the Arecibo Telescope
\citep{spi14}, the Molonglo Observatory Synthesis Telescope \citep{cal17},
and the Australian SKA Pathfinder \citep{ban17}; information on all FRBs is
listed in the FRB Catalogue \citep{pet16}.\footnote{
http://frbcat.org/}
The defining characteristic of FRBs is a dispersion measure (DM) in
excess of what can be explained by the Milky Way. If this excess is attributed to free electrons in the intergalactic medium (IGM),
FRBs could thus conceivably be used to probe the so-called ``missing'' baryon and magnetic fields and turbulence
in the IGM
(\eg \cite{rav16}). The observed rate of
FRBs is at least as high as $5\times10^3$ sky$^{-1}$
day$^{-1}$ \citep{bha18}. 

However, the origin of FRBs is still unknown. There are many theoretical proposals; for
example, mergers of double neutron star binaries \citep{tot13}, mergers of
double white dwarf binaries \citep{kas13FRB}\footnote{They assume that
the emitting region is a portion of the merged white dwarf to reproduce
the short duration of FRBs.}, the collapse of rotating supermassive neutron stars to black
holes \citep{fal14,zha14frb},
exotic Galactic compact objects \citep{kea12,ban14}, interactions between
the supernova shock and magnetosphere of a neutron star \citep{ego09},
compact objects in young supernovae \citep{con16,pir16}, supergiant
pulses from extragalactic neutron stars \citep{cor16}, and giant flares from
magnetars \citep{tho13,pen15}.

Observationally, most FRBs are non-recurrent, with the exception of
FRB~121102 \citep{spi14,spi16}. Thanks to its repeatability, radio
interferometric observations of its bursts have achieved a subarcsecond
localization and identify a faint and persistent optical counterpart
\citep{cha17}. Follow-up observations reveal that the host galaxy is an
irregular and low-metallicity dwarf galaxy at $z=0.19$ and does not show
optical signatures of AGN activity \citep{mar17,ten17} and that the
source of FRB~121102 is
spatially coincident with a star-forming region
\citep{bas17,kok17}. These observations are consistent with the proposal of newly born neutron stars
as the origin of FRBs. Unlike FRB~121102 no repeat emission has yet been
discovered in the other FRBs despite considerable effort \citep{lor07,rav15,pet15}. 

Owing to the large single-dish telescopes used to discover FRBs,
localization is typically
$\sim15$~arcmin, and there are a large number ($>10^4$) of galaxies in the localized region.
In order to clarify the origin of FRBs, the first step is a
precise localization with telescopes which have better spatial
resolution. In scenarios involving coincident optical emission this can
be achieved by using wide-field optical telescopes to monitor FRB
fields. The SUrvey
for Pulsars and Extragalactic Radio Bursts (SUPERB) operates a realtime
data analysis pipeline issuing realtime alerts \citep{kea17,bha18}. This enables searches for
contemporaneous variability in the radio, or other wavelengths, or
indeed in other windows of observation such as neutrinos or
gravitational waves. Previously \citet{kea16}
reported a variable radio source in the field of FRB~150418, which
was thought to be a fading counterpart associated with the FRB, but
continued longer-term monitoring revealed
a persistent radio source with short timescale variability
\citep{wil16,joh17}. The variability and the location are consistent
with an active galactic nuclei at the galaxy's center
\citep{aki16,bas16} but it is unclear if this is truly associated with
the FRB event, or more generally how FRBs and AGN could be associated
(see \eg \cite{vie17}).

\begin{figure*}
 \begin{center}
  \includegraphics[width=16cm]{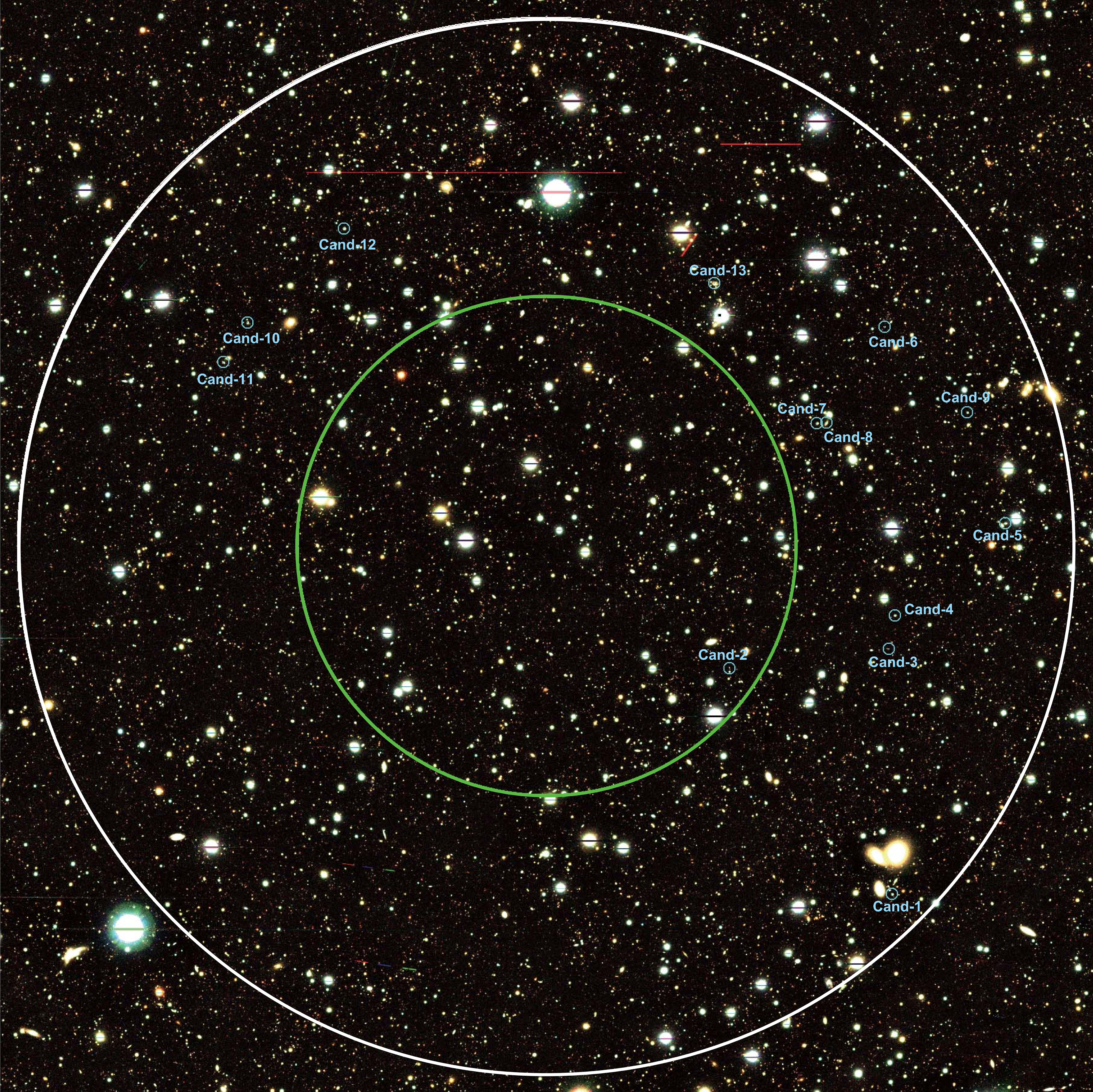} 
 \end{center}
 \caption{Three color image on Day~14 of a part of a field followed-up with HSC
 centered at the central position of the beam 04. The FWHM of beam
 profile and the localization area adopted in this paper are shown in a
 green circle with a radius of $7.05$~arcmin and a white circle with a
 radius of $15$~arcmin, respectively. The locations of 13 final
 candidates are shown in cyan circles.
  }
 \label{fig:image}
\end{figure*}

After a realtime alert from the SUPERB collaboration, we performed
optical imaging observations of \FRB\ detected in beam 04 of the
Parkes multi-beam receiver \citep{man01}. This FRB has a DM of
$960.4\pm0.5$~pc~cm$^{-3}$ corresponding to an estimated redshift of
$\sim0.8$ assuming it lies along an average line of
sight through the Universe \citep{bha18}. An optical afterglow is expected
by some theoretical models, \eg the merger of double neutron star
binaries and double white dwarf binaries. We use 
Subaru/Hyper Suprime-Cam (HSC, \cite{miy12,miy18hsc}) that is a  unique
wide-field camera on an 8m-class telescope with a field of view of
1.77~deg$^2$ and a pixel size of
$0.17$~arcsec. The HSC has the highest survey power per
unit time of any optical telescope and its field of view is 40 times larger than the typical
localization of FRBs. Thus it is the most powerful instrument for
the follow-up observations of FRB fields in the optical.

This paper consists of the following sections. In Section~\ref{sec:obs}, the
observations and data analyses are described. In
Section~\ref{sec:nature}, we present multicolor light curves of
candidates possibly associated with \FRB, and discuss their
nature. In Section~\ref{sec:constraint}, we describe constraints on the
association between \FRB\ and Type Ia supernova (SN~Ia) from the
observations. In Section \ref{sec:conclusion}, a discussion and our conclusions are
presented. In this paper, we adopt the AB magnitude system unless
otherwise noted and the fiducial cosmology with
$H_0=70~{\rm km~s^{-1}~Mpc^{-1}}$, $\Omega_\lambda=0.7$, and
$\Omega_{\rm M}=0.3$.

\section{Observations and data reduction}
\label{sec:obs}

\begin{figure}
 \begin{center}
  \includegraphics[width=8cm]{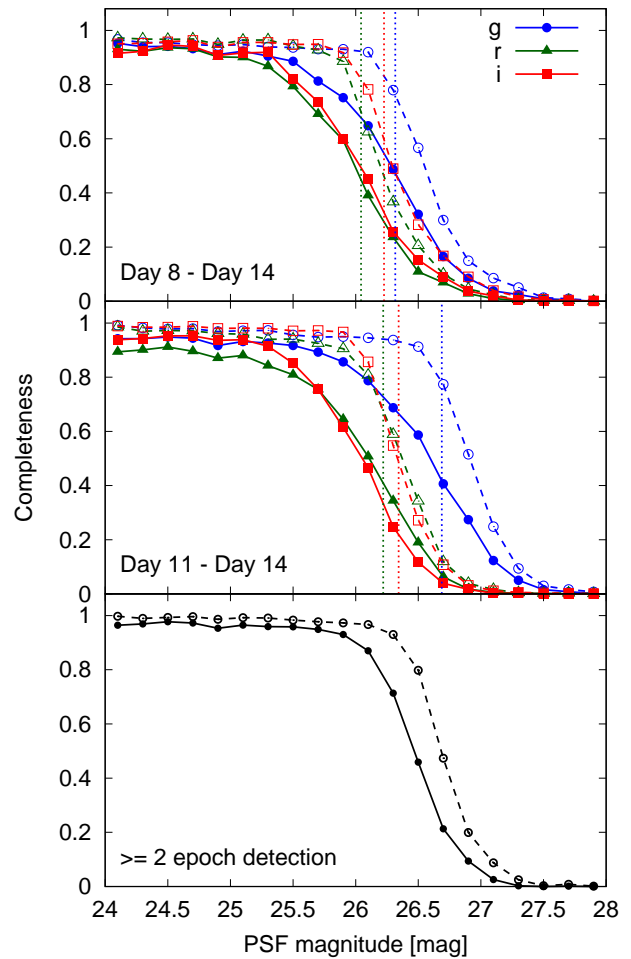} 
 \end{center}
 \caption{Completeness of source detection in the difference images
 on Day~8 $-$ 14 (top) and Day~11 $-$ 14 (middle), and in the 6 difference images
 (bottom). The dashed and solid lines represent completeness before and
 after the source screening, respectively. The vertical dotted lines
 show the $5\sigma$ limiting magnitude in the difference images.
  }
 \label{fig:completeness}
\end{figure}

\FRB\ was discovered at 2015-12-30 16:15:46 (UT) at $1.4$~GHz at \RADEC\
with an uncertainty of $\sim15$~arcmin
diameter which corresponds to the full width at half maximum (FWHM) of
the receiver beam pattern (Figure~\ref{fig:image}). In this paper, in order to not miss candidates outside the
FWHM radius, we take a radius of 15~arcmin to judge whether a candidate
is located in the localization area of \FRB. The FRB's true position is
very unlikely to be beyond this radius unless the source count slope is
unrealistically steep \citep{mac18}. Hereafter, we call this
circle the localization area of \FRB.
The mean Galactic color excess within a circle of
$5$~arcmin radius in this direction is $\Ebvg=0.039$~mag
\citep{sch11},\footnote{http://irsa.ipac.caltech.edu/applications/DUST/}
although there is a small scatter in this value of $\sim0.005$~mag across the
localization. The extinction in the host galaxies is not
corrected. We performed HSC $g$-, $r$-, and $i$-band follow-up imaging
observations with $50-70$~min exposures on Jan 7, 10, and 13, 2016 (UT,
Days 8, 11, and 14 after \FRB). Each exposure consists of multiple 3.5-
or 4-min shots with dithering. The FWHMs of the point spread
function (PSF) range from $0.7$ to $1.4$~arcsec. The observation
details are summarized in Table~\ref{tab:obs}.

The data is reduced using HSC pipeline version 4.0.5 \citep{bos17},
which is based on the LSST pipeline
\citep{ive08,axe10,jur15}. It provides packages for bias subtraction,
flat fielding, astrometry, flux calibration, mosaicing, warping,
stacking, image subtraction, source detection, and source measurement. 
The astrometric and photometric calibration is made relative to the
Pan-STARRS1 (PS1, \cite{ps1survey}) with a 4.0~arcsec (24-pixel)
aperture diameter. The images after CCD processing are warped to the predefined skymap with a
pixel size of $0.17$~arcsec and the calibrated images
are then stacked with a direct weighted average.

\begin{figure*}
 \begin{center}
  \includegraphics[width=16cm]{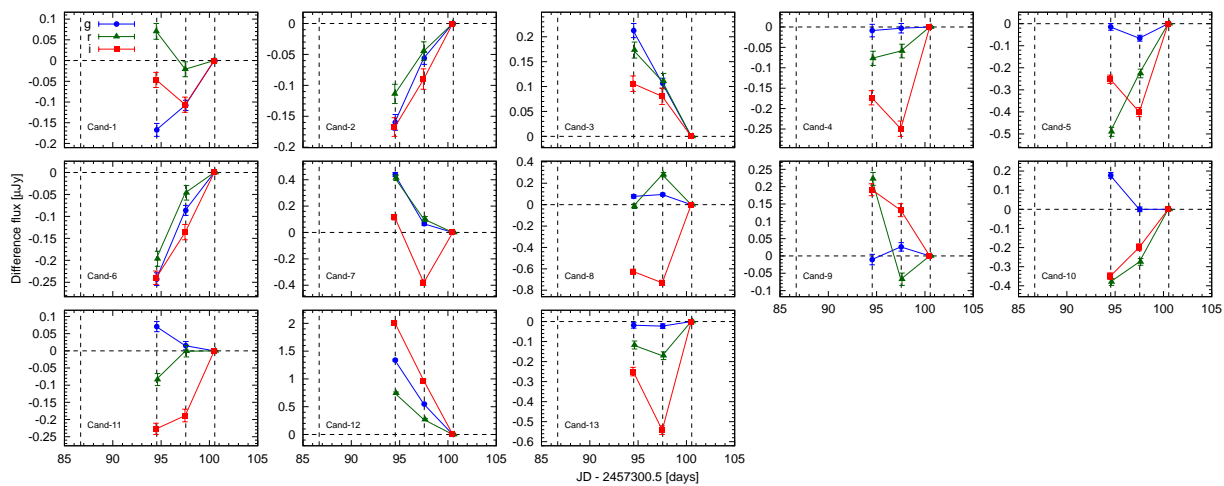} 
 \end{center}
 \caption{Difference light curves of the $13$ final candidates. The
 vertical lines represents the event time of \FRB\ and the epochs of the
 HSC observations.
 }
 \label{fig:difflcs}
\end{figure*}

We estimate the $5\sigma$ limiting magnitudes of the
$gri$ stacked images in the localization area of \FRB\ on Days $8$,
$11$, and $14$ using between $90$ and $600$ apertures with a diameter
of twice the FWHM size of PSF (Table~\ref{tab:obs}). Here, we distribute
randomly and independently the apertures on the sky area without any
detection of sources and subtract the local sky flux from the aperture
flux. The $5\sigma$ limiting magnitudes in the localization area of
\FRB\ range from $26.0$~mag to $27.3$~mag (Table~\ref{tab:obs}). The
deepest observation on Day~10 detects $\sim190,000$ sources in the
localization area of \FRB.

In order to select variable objects, we perform image subtraction
between the stacked images. The images taken on Day 14 are set as the reference
images and are subtracted from the science images taken on Days 8 and
11 because a time-variable object would typically have a larger
differential flux over a
longer time interval. The images with narrower PSF between reference and science
images are convolved with kernels to make their PSFs equivalent
\citep{ala98}. The $5\sigma$ limiting magnitudes of the difference
images in the localization area of \FRB\ are estimated with randomly
and independently distributed apertures with a diameter of twice the FWHM size of the PSF. As the image
subtraction between images with similar depths result in a shallower
depth than the original images, the $5\sigma$ limiting magnitudes of the
difference images range from $25.8$~mag to
$26.3$~mag. We also evaluate detection completeness by random
injection and detection of artificial point sources with various
magnitudes (Figure~\ref{fig:completeness}). The $5\sigma$ limiting
magnitudes are comparable to the PSF magnitude at $50-70\%$ completeness. The $5\sigma$ limiting magnitudes and the $50\%$ completeness
magnitudes in the difference images are summarized in Table~\ref{tab:subtracted}.

Point sources in the difference images are detected and measured with
the HSC pipeline. Since there are many false positives due to noise, we
adopt the following criteria to identify bona fide sources;
(1) significance higher than $5\sigma$, (2)
$(b/a)/(b/a)_{\rm PSF} >0.65$ where $a$ and $b$ are the lengths of the
major and minor axes of a shape of a source, respectively, (3) FWHM of
a source is between 0.7 to 1.3 times that of the PSF, (4) the residual
of PSF subtraction from a source is $<3\sigma$. These
criteria are similar to those adopted in searches for optical
counterparts of gravitational wave signals
\citep{uts17gw151226,tom17gw170817}. After this screening, 118
and 1078 variable sources remain inside and outside the localization area of \FRB,
respectively.
We also estimate the detection
completeness after the screening with injected artificial
sources. The $50\%$ completeness magnitudes after the 
screening are $0.2$-$0.4$~mag shallower than those before owing to low
signal-to-noise ratios (S/N). We further imposed the sources be detected at least twice at
the same location in the 6 difference images. The $50\%$
completeness magnitude after this additional screening is $26.5$~mag for
point sources, which
is the deepest among all the optical follow-up observations of FRBs
which cover the entire localization area
\citep{pet15frb140514,kea16,bha18}.

The PSF shape in the stacked images sometimes fails the image
subtraction at the center of galaxies leaving an unphysical
residual. Thus, we also perform an alternative procedure of image
subtraction; the stacked reference image is subtracted from each
exposure of the science images and the resultant difference images are
stacked. If the source is located within a separation of $0.34$~arcsec
($2$~pixel) from objects in
the stacked images or objects in the PS1 catalog, we check whether the
source is detected in the alternative procedure. Here, the positions of
objects in the stacked images are measured with the HSC pipeline at the
epoch when the source is faintest. We exclude the source if
it is not detected in the alternative procedure. After this exclusion,
53 variable sources inside the
localization area of \FRB\ and 490 variable sources outside the localization area
of \FRB\ remain,
respectively. However, false positives, \eg failures of image
subtraction around bright objects (\eg Figure~1 in
\cite{morii16}), still remain. Therefore, we visually inspect the variable
sources to remove these, and select 13 sources inside the localization
area of \FRB\ as the final candidates. These are
summarized in Table~\ref{tab:cand}. Transients outside the localization
area of \FRB\ are described in Appendix~\ref{sec:outside}.

Difference light curves of candidates are derived with 
forced PSF photometry of the difference images with aperture correction
at the location of candidates and dereddening for the Galactic
extinction. Figure~\ref{fig:difflcs} shows their difference light curves. We
also measure the PSF fluxes at the location of candidates in the stacked
images with forced PSF photometry. The forced PSF flux in the stacked image is
an upper limit of the flux of the candidate because it consists of the
fluxes of both the candidate and its host galaxy.

\begin{figure*}
 \begin{center}
  \includegraphics[width=16cm]{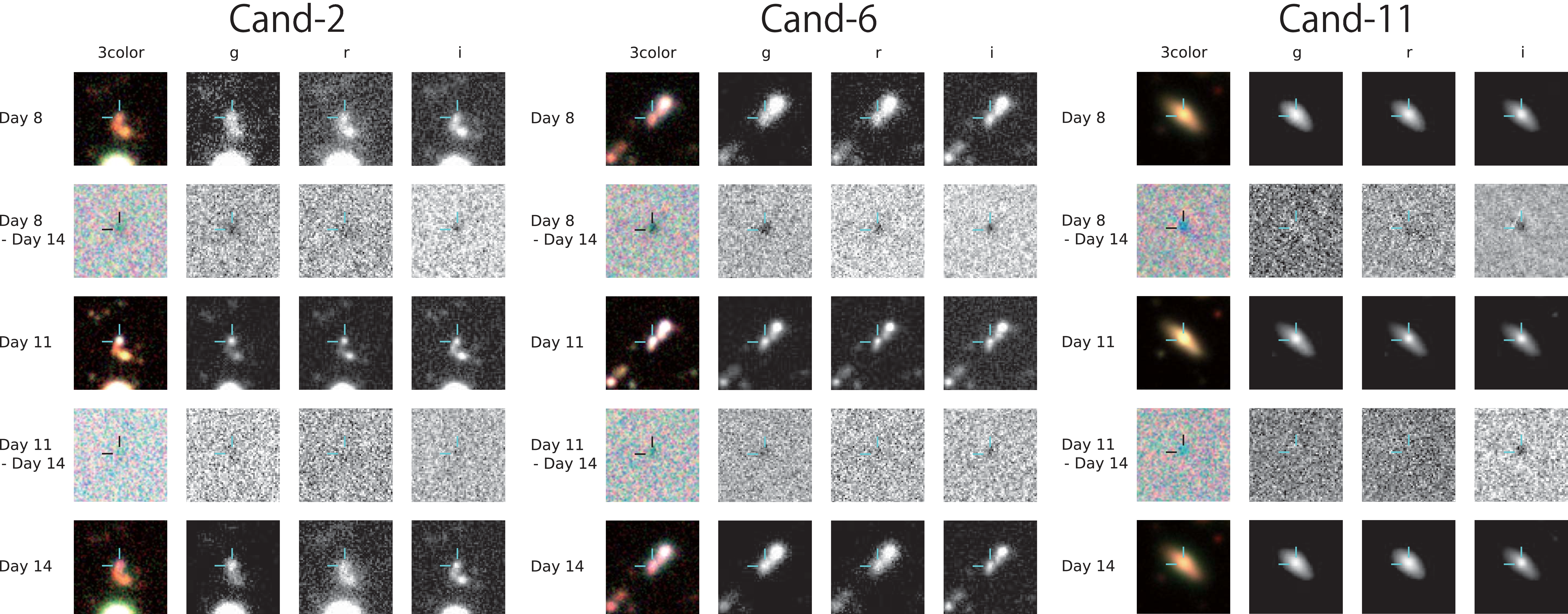} 
 \end{center}
 \caption{Images at the location of the candidates fitted with the transient
 templates (ticks). The lengths of ticks
 are $1$~arcsec and the figure size is $10\times10$~arcsec$^2$. 
}
\label{fig:cutout}
\end{figure*}

\begin{figure*}
 \begin{center}
  \includegraphics[width=16cm]{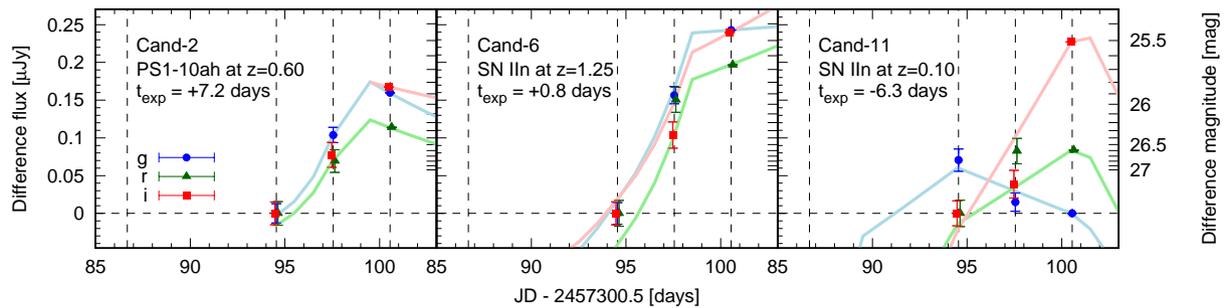} 
 \end{center}
 \caption{Comparisons of the difference light curves of the candidates and the best-fit
 templates. The fluxes are shifted to be zero at the faintest epoch.
}
\label{fig:LCsDate}
\end{figure*}

\section{Nature of candidates}
\label{sec:nature}

After the source screening and the visual inspection, 13 final
candidates remain. Their locations are shown in Figure~\ref{fig:image}.

\subsection{Light curve fitting and locations within galaxies}

In order to investigate the nature of the candidates with the limited
available data over $\sim1$~week, we primarily perform the difference
light curve fitting described in Appendix~\ref{sec:LCfit}.
For this, we adopt templates of SNe~Ia
(\cite{hsi07}), core-collapse SNe (CCSNe; Type Ibc, IIP, IIL, and IIn
SNe, \cite{nug02}), and rapid transients (RTs, from the gold sample in
\cite{dro14}).\footnote{We exclude PS1-12bb from the sample because it
does not have the data points before the peak.} We take into account a
variation of the explosion date
$\texp$ of templates since the occurrence time of \FRB\ and an intrinsic
variation of templates, \eg peak absolute magnitude and extinction in
the host galaxy. In order to judge the agreement of the template
fitting, we employ $Q(\chi^2|\nu)$ value, the probability that a $\chi^2$
distribution with a
degree of freedom $\nu$ exceeds a given $\chi^2$ value by chance.

In addition to the difference light curve fitting, if the flux of the host galaxy is
likely to dominate the flux of the candidate, we adopt the photometric
redshift of the host galaxy as a criterion to narrow down the possible templates and impose
that the redshift derived from the difference light curve fitting be
consistent with the photometric redshift. The
photometric redshift is derived with the $gri$ photometry of the host
galaxy in the stacked images using a photometric redshift code with
Bayesian priors on physical properties of galaxies
\citep{tanaka15photoz}. Further, a separation of each candidate from
its host galaxy (if existing) is employed to exclude the
possibility of active galactic nuclei (AGN) and Galactic stars. We
conservatively adopt both the position measured with the HSC pipeline and the
position of a pixel with a peak count in a $10$~pix$\times10$~pix square
centered on the candidate as the possible position of the associated object
in the stacked images at the epoch when each candidate is faintest.

In this subsection, we
focus on candidates that can be distinguished from optical variability
typical of AGNs. As any optical counterpart to \FRB\ need not have a
light curve of known form, we consider (1) candidates reproduced with transient templates
with a loose criterion $Q>10^{-4}$ because our aim is only to narrow down possible
templates, or (2) candidates located off-center of their host galaxy with a
separation of $>0.34$~arcsec. 
After applying the above steps, we pick up 4 candidates
(Table~\ref{tab:model}). Nine of the 13 candidates are consistent with
optical variability from AGN, and we discard them from further
investigation and focus on these 4 candidates in this subsection.
Of the remaining 4, 3 candidates are well fitted by the
transient templates with $Q>10^{-4}$. 

\subsubsection{Candidates consistent with transient templates}

Three candidates, \SNa, \SNc, and \SNd, are well
fitted with the transient templates. These candidates are
associated with extended objects, presumably galaxies
(Figure~\ref{fig:cutout}). Their separations from the center of the
localization area of \FRB\ are $6.24$~arcmin for \SNa, $11.43$~arcmin
for \SNc, and $10.55$~arcmin for \SNd. As the fluxes in the stacked images are
dominated by the fluxes of their host galaxies (Table~\ref{tab:obs}), we
derive the photometric redshifts of the host galaxies adopting the
fluxes on Day 8, on which the contributions of the 3 candidates to the
total fluxes are relatively small. The photometric redshifts of the host
galaxies of \SNa, \SNc, and \SNd\ are $z=0.18^{+0.46}_{-0.15}$,
$1.44^{+0.46}_{-0.27}$, and $0.19^{+0.07}_{-0.15}$, respectively. Here,
we adopt the $95\%$ probability as the errors.
The photometric redshift of the host galaxy of \SNc\ is inconsistent
with the maximum redshift inferred from the DM of \FRB\ and thus \SNc\
is unlikely to be an optical counterpart of \FRB. However, in the next
paragraph, we also
describe the templates consistent with \SNc\ for comparisons with a
theoretical estimate and the total
number of candidates outside the localization area.

The best-fit templates within the $95\%$ probability of the photometric
redshift are an RT (PS1-10ah) at $z=0.60$ for \SNa, an SN~IIn at $z=1.25$ for
\SNc, and an SN~IIn at $z=0.10$ for \SNd\ (Figure~\ref{fig:LCsDate}). We also
derive probable templates and possible ranges of parameters
allowing the template with the $Q$ value of $>0.01$ of the
maximum $Q$ value for each candidate. While \SNc\ and \SNd\ can be fitted only
with the SN IIn template, \SNa\ can also be fitted with SN IIP and
IIn templates in addition to the best-fit RT template
(Table~\ref{tab:model}).  As the difference of the $Q$ values of \SNa\
between the RT and SN~IIn templates is only a factor of $2$, an SN~IIn
is a comparably good fit to an RT for the origin of \SNa.
The location of the candidate can be used to constrain the possibility
of an AGN origin. While \SNa\ is located
off-center of the galaxy, \SNd\ is located at the
center of the host galaxy and thus optical variability of an AGN cannot
be excluded.
The most probable templates for \SNa\ include those with $\texp\sim0$~day. If \SNa\ exploded at
the same date as \FRB, the achromatic and rapid rising of \SNa\ is
only consistent with SNe~IIn, while the RT templates agree with
the multicolor light curves of \SNa\ only for
$\texp=+7.2^{+1.1}_{-0.5}$~days.
 
\begin{figure}
 \begin{center}
  \includegraphics[width=8cm]{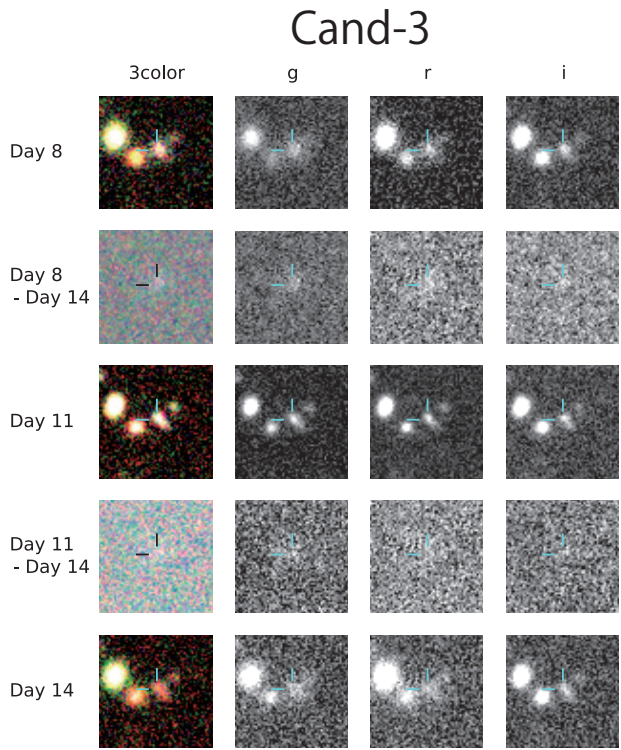} 
 \end{center}
 \caption{Images at the location of \SNb\ (ticks) that is located at the
 off-center of extended objects. The lengths of ticks
 are $1$~arcsec and the figure size is $10\times10$~arcsec$^2$. 
}
\label{fig:cutoutNote}
\end{figure}

 \subsubsection{A candidate inconsistent with transient templates}
\label{sec:offcenter}

 The multicolor light curves of an off-center candidate, \SNb, cannot be
 reproduced with any of our transient templates.
 \SNb\ is located $0.43$~arcsec away from the center of its host
 galaxy, and $10.15$~arcmin from the center of the localization area
 of \FRB\ (Figure~\ref{fig:cutoutNote}). Its off-center nature
 indicates that \SNb\ is likely not due to optical variability of
 an AGN. 
 As the contribution of \SNb\ to the total flux in the stacked
image is not negligible, we can not evaluate the photometric redshift of
 the host galaxy.

\begin{figure}
 \begin{center}
  \includegraphics[width=8cm]{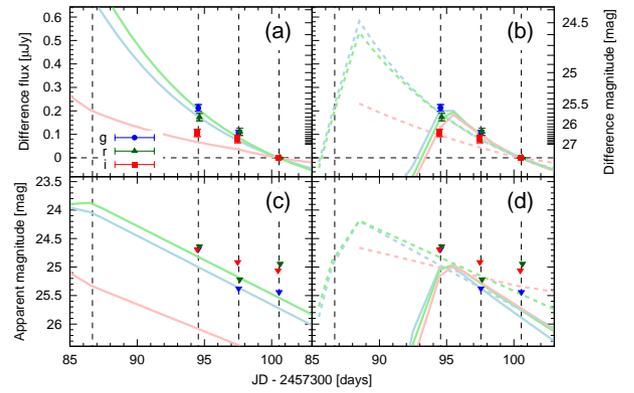} 
 \end{center}
 \caption{(ab) Comparisons of the difference light curves of \SNb\
 and the best-fit RT templates with $s=1$ [PS1-11qr, (a)], $s=0.8$
 [PS1-10ah, dashed in (b)], and $s=0.6$ [PS1-10bjp, solid in (b)]. The fluxes are shifted to be
 zero at the faintest epoch. (cd) Comparisons of the upper limits
 for the fluxes of \SNb\ in the stacked images and the apparent
 magnitudes of the best-fit RT templates with $s=1$ [PS1-11qr, (c)], $s=0.8$
 [PS1-10ah, dashed in (d)], and $s=0.6$ [PS1-10bjp, solid in (d)].
}
\label{fig:LCsSNb}
\end{figure}

The best-fit template is an RT template (PS1-11qr) being $+0.9$~mag
fainter than PS1-11qr and having $Q=2.2\times10^{-6}$
(Table~\ref{tab:model}). The low $Q$ value stems mainly from the poor
agreement in the $i$-band difference light curve
(Figure~\ref{fig:LCsSNb}a). The most probable template other than
PS1-11qr is also an RT template (PS1-12brf) being $+0.45$~mag fainter
than PS1-12brf with $Q=2.3\times10^{-8}$. On the other hand, comparisons
with SN~Ia and CCSN templates result in a $Q$ value of $<10^{-21}$.

The forced PSF magnitudes of \SNb\ in the stacked images indicate a
$g$-band decline rate of $>+0.23$~mag~day$^{-1}$ in the observer frame. The rate is
consistent with the rest-frame decline of the most rapidly-declining
RTs in \citet{dro14}. Although this could be consistent with
kilonova emission from binary neutron stars coalescence
\citep{utsumi17}, the blue color of \SNb\ on Day~8 is inconsistent with
that regardless of the occurrence date
because a kilonova with lanthanoid production is red in optical
$0.5$~days after the coalescence and becomes redder with time
(\eg \cite{pian17}). These results demonstrate that the
multicolor light curves of \SNb\ can be reproduced only by the RT
templates. We also note that, if we set the criterion to be
$Q>2\times10^{-6}$, there are no candidates fitted with the RT templates
other than \SNa\ and \SNb, and Cand-4 is fitted with
an SN~IIL template with $Q=5.8\times10^{-5}$.

The reason why \SNb\ is not reproduced with the transient templates is the
small fluxes in the stacked images, which are faint especially on Day 11
and 14. This could be resolved if \SNb\ has fainter peak magnitude and/or more rapid decline than the
RTs observed so far. Since the distribution of the peak magnitude and
decline rate of RTs is still not well-known, we test the difference light
curve fitting stretching a timescale parameter of the RT templates.
Here, we reduce the stretch factor $s$ from $1.0$ to $0.05$ in steps
of $0.05$. Doing this, agreement between \SNb\ and RT templates with
$Q>10^{-4}$ can be realized in the range
$0.15\leq s\leq0.8$ with the best agreement achieved with $s=0.6$. The
best-fit templates with $s=0.8$ and $0.6$ are shown in
Figures~\ref{fig:LCsSNb}(b) and \ref{fig:LCsSNb}(d), and the probable templates with $s=0.8$ and
$0.6$ are summarized in Table~\ref{tab:modelSNb}. For $s=0.8$, the best-fit template is
the PS1-10ah template at $z=0.2$, which is
$+1.0$~mag fainter than PS1-10ah, and has $Q=7.4\times10^{-4}$. The possible ranges of $\texp$ and
$z$ are $\texp=-1.4^{+7.5}_{-2.5}$~days and $z=0.20^{+0.68}_{-0.03}$, respectively. 
For $s=0.6$, the best-fit template is the PS1-10bjp template at $z=0.4$,
$+1.0$~mag fainter than PS1-10bjp, and with $Q=4.2\times10^{-1}$. The possible ranges of $\texp$ and $z$ are
$\texp=+5.2^{+1.5}_{-7.5}$~days and $z=0.40^{+0.13}_{-0.08}$, respectively. The same
explosion date as \FRB\ is included in both of the possible templates
with $s=0.8$ and $0.6$.

\subsection{Multiwavelengh observations and host galaxies}
\label{sec:associatedobject}

In addition to the multicolor light curves of candidates and the
candidate locations in the host galaxies, the multi-wavelength
observations and the properties of the host galaxies are useful to
constrain the nature of candidates.

Assuming that \FRB\ is located at the locations of candidates, we
estimate the intrinsic radio S/N of \FRB\ and summarize it in
Table~\ref{tab:cand}. Taking into account these locations and the
putative intrinsic brightness, \FRB\ at the location of \SNg\ can be
detected with beam 01 of the receiver at Parkes with an S/N of $\sim7$. On the
other hand, \FRB\ at the location of the other candidates cannot be 
detected with the other beams. The non-detection of \FRB\ in the other
beams excludes \SNg\ as an optical counterpart to \FRB.
Here, we assume a Gaussian profile for each feed horn of the multi-beam
receiver at Parkes \citep{man01}.

\citet{bha18} reported several radio sources detected in the follow-up
observations with the Australia Telescope Compact Array (ATCA,
\cite{ATCA}), the Karl G. Jansky Very Large Array (VLA), and the Giant
Metrewave Radio Telescope (GMRT, \cite{GMRT}). None of these is
located within two times the FWHM of the synthesis beams from the sources
detected by VLA and GMRT. On the other hand, Cand-1 and Cand-10 (likely
AGN) are located at a
distance of about twice the FWHM of the synthesis beam at $5.5$~GHz from the
sources detected only with ATCA although any association cannot be 
conclusive because there are other optical sources in the wide elongated
synthesis beam of ATCA. Our other candidates are not associated with
any radio sources. The radio flux densities at $5.5$~GHz and $7.5$~GHz are
$22.2\pm0.1$~mJy and $20.2\pm0.1$~mJy for Cand-1 and $0.84\pm0.06$~mJy
and $0.62\pm0.07$~mJy for Cand-10. The spectral slopes are
$\alpha_{\rm LF}=-0.31\pm0.03$ and $-0.96\pm0.57$ for Cand-1 and Cand-10,
respectively. These slopes are consistent with $\alpha_{\rm LF}$ of AGN
\citep{planck16radioagn}. Additionally, the optical light curves of
Cand-1 and Cand-10 are inconsistent with the transient
templates. Therefore, if Cand-1 and Cand-10 are associated with the
radio sources reported by \citet{bha18}, they are likely to be optical variabilities of
AGN. The persistent optical counterparts of the radio sources are described in
Appendix~\ref{sec:radio}.

We further compare the candidate locations with the Wide-field Infrared
Survey Explorer (WISE) catalog \citep{NEOWISEcatalog}, and find that
only the host galaxy of \SNa\ could host a AGN based on the color of
$W1-W2\geq0.8$~mag (Vega, \cite{ste12}). However, there are multiple 
galaxies within a resolution element of WISE (Figure~\ref{fig:cutout}) and
\SNa\ is located off-center of the host galaxy. Thus, \SNa\ 
is not necessarily the optical variability of an AGN. Furthermore, we
check the ROSAT catalog \citep{ROSATcatalog}, the 3XMM-DR7 catalog
\citep{XMMcatalog}, and the NRAO VLA Sky Survey (NVSS, $1.4$~GHz)
catalog \citep{NVSScatalog} but there are no associated sources with our
13 variable candidates in these catalogs.

Along with the photometric redshifts of the host galaxies, their stellar
masses and star formation rates (SFRs) are also derived for the $12$
candidates other than \SNb. The photometric redshifts of the host
galaxies of the $11$ candidates other than \SNb\ and \SNc\ are consistent with the
maximum redshift inferred from the DM of \FRB. While the stellar
masses of the host galaxies of the $11$ candidates other than \SNa\ and
\SNb\ are
$\geq10^{9}~\Msun$, the stellar mass of the host galaxy of
\SNa\ is $8.6\times10^7~\Msun$ comparable with the stellar mass of the
host galaxy of FRB~121102 \citep{bas17,kok17,ten17}. On the other hand, the
SFR of the host galaxy of \SNa\ is $0.03~\Msun$~yr$^{-1}$ an order
of magnitude lower than that of the host galaxy of FRB~121102. The
host galaxy of \SNa\ is classified as a star-forming galaxy and it is
consistent with the conclusion that \SNa\ is CCSN or RT. The
host galaxies of the $11$ candidates other than \SNb\ and Cand-5 have the SFR of
$>10^{-2}~\Msun$~yr$^{-1}$ as a median value and are consistent with hosting CCSNe.

\begin{figure}
 \begin{center}
  \includegraphics[width=8cm]{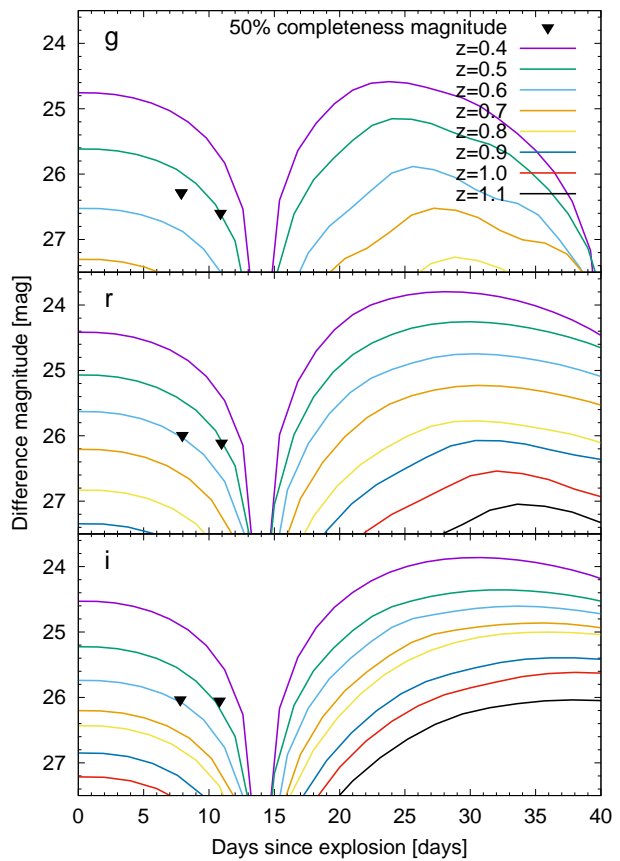} 
 \end{center}
 \caption{Magnitude of absolute difference fluxes of SN Ia templates with respect to
 Day~14. The $50\%$ completeness magnitudes are shown with downward
 triangles. 
 }
 \label{fig:LCday14}
\end{figure}

 \section{Testing Type Ia Supernovae as the origin of \FRB}
\label{sec:constraint}

The comparisons of the 13 candidates with SN~Ia templates give $Q<4\times10^{-8}$. Thus, we conclude that all of
candidates are inconsistent with SN~Ia templates.
In this section, we illustrate how the non-detection constrains the
association of \FRB\ with an SN~Ia. We adopt the light curve of SNe~Ia exploded
at Day~0, \ie $\texp=0$, with $s=1$ and $c=0$.

Figure~\ref{fig:LCday14} shows the difference magnitude with respect to
the flux at Day~14. The PSF magnitudes with detection completeness of
$50\%$ are also shown. The difference magnitude of SN~Ia is brighter for longer
time intervals. The $50\%$ completeness magnitude is comparable with the
difference magnitude of SN~Ia at $z\sim0.6$ between Days~8 and
14. Therefore, our
analysis can exclude the association of SNe~Ia with $\texp=0$ at
$z\leq0.6$ with \FRB. On the other hand, the DM of \FRB\
puts an upper limit on its redshift of $z\sim0.8$ assuming
an average line of sight in the Universe \citep{bha18}. Our
results leave room for SNe~Ia association with \FRB\ at
$z\sim0.6-0.8$. In this range, this constraint then corresponds to an upper limit of
the DM of the host galaxy of \FRB. If an SN~Ia is associated with \FRB,
the DM of the host galaxy of \FRB\ needs to be less than
$\sim300$~pc~cm$^{-3}$ \citep{mcq14}.

\section{Discussion \& Conclusion}
\label{sec:conclusion}

We performed optical follow-up observations of \FRB\ using Subaru/HSC on
Days~8, 10, and 14 after a realtime alert from the SUPERB
collaboration \citep{bha18}. The survey field covers the localization area of \FRB,
and the $50\%$ completeness magnitude is $26.5$~mag for point sources, which is
the deepest among the optical follow-up observations of FRBs covering
the entire localization area.

\begin{figure*}
 \begin{center}
  \includegraphics[width=16cm]{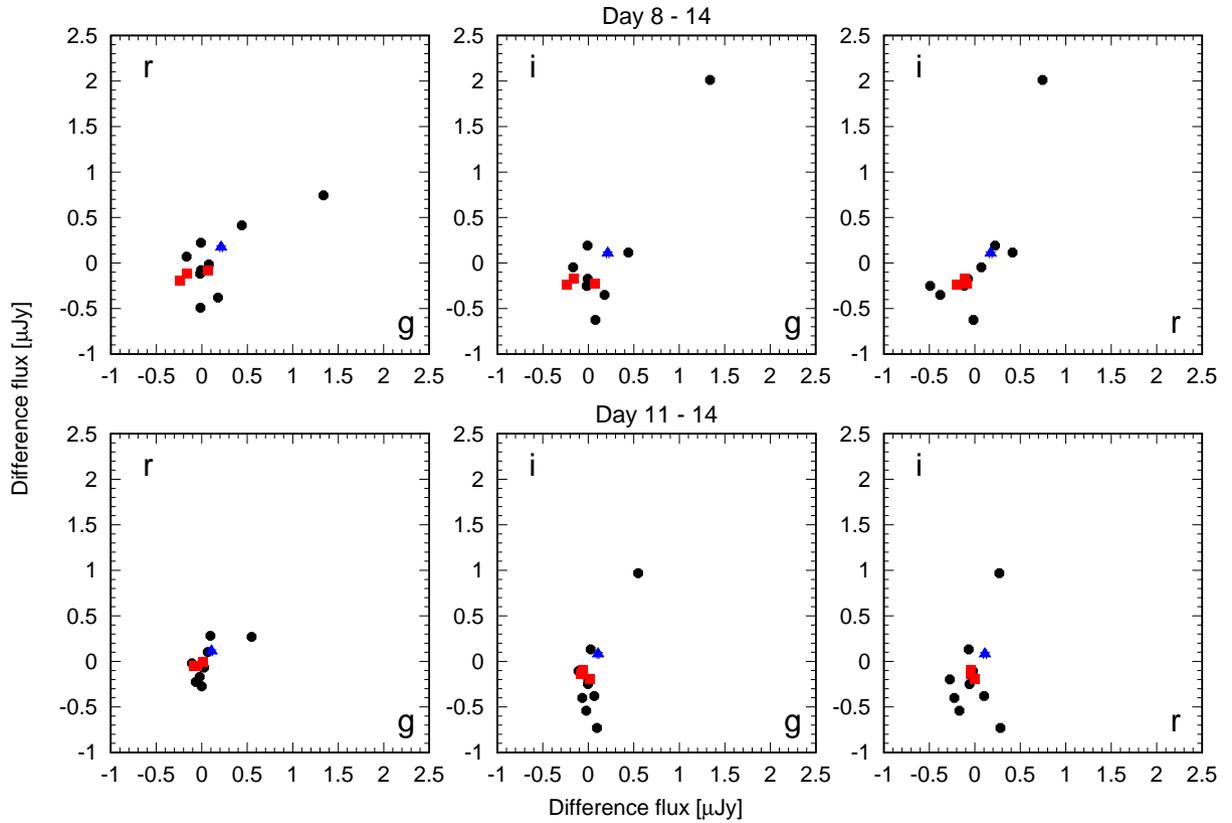} 
 \end{center}
 \caption{$g$-, $r$-, and $i$-band difference fluxes of
 candidates (red: \SNa, \SNc, and \SNd,
 blue: \SNb, and black: the others). 
}
\label{fig:diffLCs}
\end{figure*}

After various sifting and screening techniques are applied and
subsequent visual inspection, we find $13$ candidates in the
localization area of \FRB. The properties of $13$ candidates are
summarized in Table~\ref{tab:scores}. Among them, the association of Cand-5 with
\FRB\ is excluded by the non-detection in the radio of \FRB\ in the other
beams at Parkes. Another 8 candidates are inconsistent with the transient templates
and are located at the center of extended objects, and thus could be optical
variabilities of AGNs.  In particular, Cand-1 and Cand-10 are
located at a distance of about twice the FWHM from the radio sources
detected with ATCA, and the spectral slopes of the radio sources are
consistent with that of the AGN. Although the possibility that one of
these 8 candidates and the optical variability of AGN are associated with
\FRB\ is not ruled out, it is also possible that all of them are
unrelated to \FRB\ and the association between them and \FRB\ can not be
investigated from our optical follow-up observations.
The photometric redshifts of the host
galaxies of these 8 candidates are consistent with the maximum redshift
inferred from the DM of \FRB.

After considering the agreement of light curves fit with known
transient templates, the location of candidates in host galaxies,
and photometric redshifts, we focus on $3$ candidates. The final
candidate, \SNc, is associated with a galaxy with a photometric
redshift higher than the maximum redshift estimated from the DM of
\FRB. They are not prominent in $g$-, $r$-, and
$i$-band difference fluxes (Figure~\ref{fig:diffLCs}). Two
candidates among the $3$ that we deemed to be most interesting are well fitted with
transient templates with $Q>10^{-4}$. Their most probable templates are RTs at $z=0.60^{+0.03}_{-0.23}$,
SN~IIP at $z=0.20^{+0.03}_{-0.03}$, and SNe~IIn at
$z=0.55^{+0.08}_{-0.23}$ for \SNa, and SNe~IIn at
$z=0.10^{+0.03}_{-0.03}$ for \SNd.
\SNd\ is located at the center of an extended object and thus could be
due to the optical variabilities of AGN. 
\SNb\ is not well fitted with the
transient templates and located off-center of its
host galaxy. However, by stretching the RT templates, the multicolor light
curves of \SNb\ are reproduced with $0.15\leq s \leq0.8$. This
suggests that \SNb\ is an RT with a rather faint peak and rapid
decline. \SNa\ and \SNb\ could have exploded at the same date
as \FRB, although some theoretical models for FRBs do not necessarily
require this. However, we can not conclude their association with \FRB.

According to a theoretical estimate with a mock catalog of SNe and
observed SN rates \citep{nii14frb}, the
number of candidates reproduced with SN templates, \ie \SNa,
\SNc, and \SNd, is consistent with the total SN rate, including SNe~Ia and CCSNe. Also, the number density of candidates
reproduced with the transient templates inside the
localization area of \FRB\ is roughly consistent with that 
outside the localization area of \FRB\ (Appendix~\ref{sec:outside}) and
that of SNe detected in a previous study with a time interval of
$6$~days \citep{mor08variable}.

No candidates are reproduced with the SN~Ia template. Comparing the $50\%$ completeness magnitude with the
difference light curve of SN~Ia templates, our observation is sensitive
enough to detect SNe~Ia with $\texp=0$ at $z\leq0.6$. According to the DM
of \FRB, this leaves room for an SNe~Ia association with \FRB\ only at
$z\sim0.6-0.8$. In other words, if an SN~Ia is associated with \FRB,
the DM of the host galaxy of \FRB\ needs to be less than
$\sim300$~pc~cm$^{-3}$.

\SNa\ and \SNb\ can be well fitted as RTs, of which the volumetric rate is
$4-7\%$ of the CCSN rate \citep{dro14}. 
While \SNa\ might be an SN~IIn according to the $Q$ values, the
multicolor light curves of \SNb\ can be reproduced only with
RTs with a faint peak and rapid decline. The light curve of \SNb\ is well
reproduced with the stretched RT templates with $0.15\leq s \leq0.8$ with $Q>10^{-4}$. 
The most probable templates for \SNb\
include the RT templates with the same explosion date as \FRB.
We compute the expected number of RTs which
will be detected coincidentally in the localization area with $g$-band
differential flux $> 0.2 \upmu$Jy in a six-day interval using the method
outlined in \citet{nii14frb}.
We utilize
the best-fit template for \SNb\ after stretching, the resultant number of coincident RTs in the
localization area is as small as $0.038$, and the probability to have
this event during the observation is thus $3.6\%$ according to the
Poisson distribution.
Outside the localization
area of \FRB, the number of candidates best-fitted with the RT template
is 1. These results suggest that the coincident
detection of an RT irrelevant to \FRB\
is relatively unlikely in the localization area. 
Thus, if \SNb\
is an RT, it may relate to
\FRB. The volumetric rates of RTs and FRBs are
consistent \citep{dro14,kea15,bha18}.

There are two possible mechanisms in the literature for an RT
and FRB to be associated; (1) the RT emission from an
ultra-stripped Type Ic supernova from a close interacting binary system
\citep{suw15,tau15,moriya17,yos17} and the FRB emission from the interaction
between supernova shock and magnetosphere of a neutron star
\citep{ego09}, and (2) the RT emission from an accretion-induced collapse of
the merger remnant of He and CO white dwarfs \citep{bro17} and the FRB
emission from the collapse of a strongly magnetized supermassive
rotating neutron star to a black hole immediately after the
accretion-induced collapse \citep{fal14,zha14frb,moriya16}. However,
there is a major caveat that the RT ejecta can be optically thick at radio
frequencies immediately after the explosion and can totally absorb the
FRB emission \citep{con16,pir16,kas17}. 

Our observation demonstrates that deep and wide observations detect
several unassociated transients in the localization area of FRB. Thus, it is
important to establish a method to exclude them. Theortical studies need
to more precisely predict light curve evolution to identify the optical
counterparts of FRBs.

The optical observations give a lower limit on the redshift. This is complementary with the upper
limit on redshift of FRBs that can be estimated by the DM of FRBs. We
encourage rapid reporting of FRBs, especially those
with low DM values, to enable future wide-field optical work aiming at
identifying any and all associated emission in this band.

\begin{ack}
This research has been supported in part by the research grant program
of Toyota foundation (D11-R-0830), the
natural science grant of the Mitsubishi Foundation, the research grant
of the Yamada Science Foundation, World Premier
 International Research Center Initiative, MEXT, Japan, and by
 JSPS KAKENHI Grant Numbers JP15H02075, JP15H05440,
 JP15K05018, JP16H02158, JP17H06362, JP17H06363, JP17K14255,
 JP18K03692. EP receives funding from the European Research
 Council under the European Union Seventh Framework Programme
 (FP/2007-2013) / ERC Grant Agreement n. 617199.
We thank the LSST Project for making their code available as
free software at http://dm.lsstcorp.org.
Funding for SDSS-III has been provided by the Alfred P. Sloan
Foundation, the Participating Institutions, the National Science
Foundation, and the U.S. Department of Energy Office of Science. The
SDSS-III web site is http://www.sdss3.org/.
SDSS-III is managed by the Astrophysical Research Consortium for the
Participating Institutions of the SDSS-III Collaboration including the
University of Arizona, the Brazilian Participation Group, Brookhaven
National Laboratory, Carnegie Mellon University, University of Florida,
the French Participation Group, the German Participation Group, Harvard
University, the Instituto de Astrofisica de Canarias, the Michigan
State/Notre Dame/JINA Participation Group, Johns Hopkins University,
Lawrence Berkeley National Laboratory, Max Planck Institute for
Astrophysics, Max Planck Institute for Extraterrestrial Physics, New
Mexico State University, New York University, Ohio State University,
Pennsylvania State University, University of Portsmouth, Princeton
University, the Spanish Participation Group, University of Tokyo,
University of Utah, Vanderbilt University, University of Virginia,
University of Washington, and Yale University. 
\end{ack}

\appendix

\begin{figure*}
 \begin{center}
  \includegraphics[width=16cm]{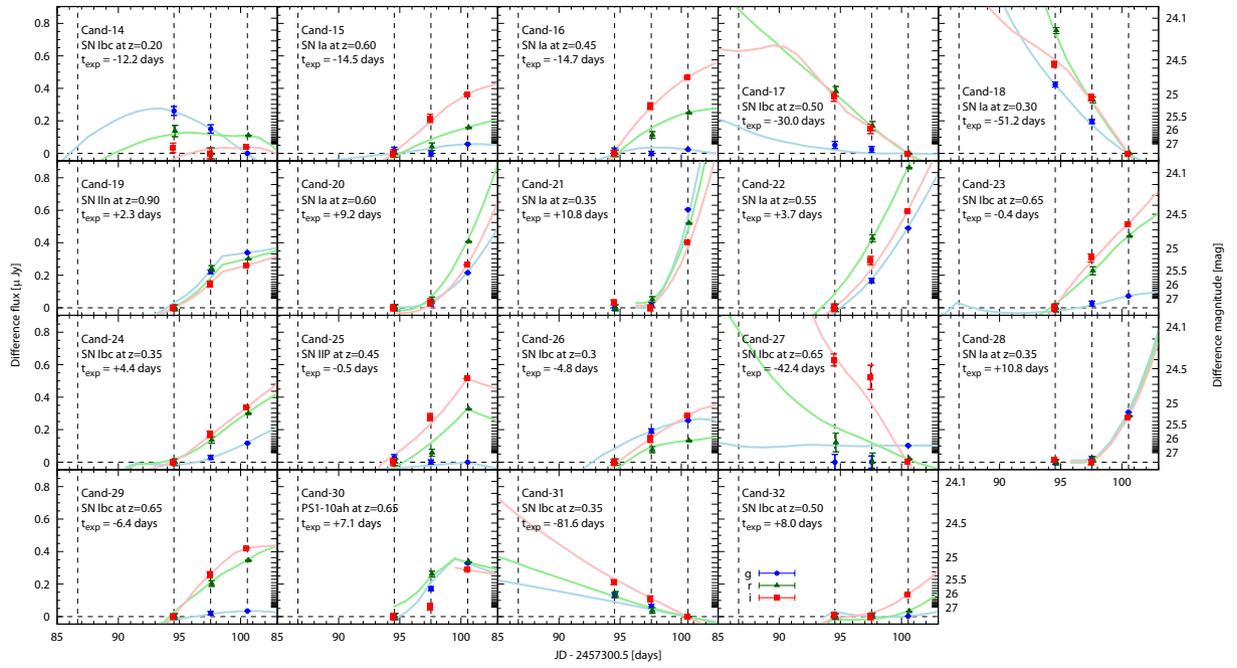} 
 \end{center}
 \caption{Comparisons of the difference light curves of the transients
 outside the localization area of \FRB\ and the best-fit
 templates. The fluxes are shifted to be zero at the faintest epoch.
 }
 \label{fig:LCout}
\end{figure*}

 \section{Transients outside the localization area of \FRB}
\label{sec:outside}

As the field of view of HSC is 9 times wider than a circle with a radius
of $15$~arcmin, \ie the localization area of
\FRB\ adopted in this paper, although the depth
outside the localization area is $0.34$~mag shallower than that inside the
localization area. Thus, we have a large sample of candidates being
irrelevant to \FRB, which can be a control sample. 
We perform the visual inspection and the light curve fitting for the 490
candidates outside the localization area of \FRB. If the photometric redshift
of the host galaxy is available, only the templates in the $95\%$ probability
are allowed. As a result, 19 visually-selected candidates\footnote{The
number of candidates reproduced with the transient templates is $17$ if
we adopt the criterion of $Q>10^{-4}$.} are fitted well with the transient templates (Table~\ref{tab:model3} and
Figure~\ref{fig:LCout}) with the $Q$ values better than the best-fit
template for \SNb, \ie $Q>2\times10^{-6}$. Taking into account the wide area and the
shallow depth outside the localization area
of \FRB, the number density of candidates reproduced with the transient templates is
roughly consistent with that of candidates inside the localization area of \FRB.
The best-fit templates are SNe~Ia for 7 candidates, CCSNe for 11
candidates, and an RT for 1 candidate. Considering the small number
statistics, the ratio of the numbers of the
candidates consistent with RT
templates to those with CCSN templates is consistent
with the ratio of RTs to CCSNe ($4\%-7\%$, \cite{dro14}).

\section{Light curve fitting}
\label{sec:LCfit}

For the light curve fitting, we adopt templates of SNe~Ia
(\cite{hsi07}), CCSNe (\cite{nug02}), and RTs (\cite{dro14}). In order
to take into account intrinsic variations, we
allow variations of several parameters; stretch $s$, color $c$, and
intrinsic variation $I$ for the SN~Ia templates, and peak absolute
$B$-band magnitude $\MB$ and extinction in the host
galaxy for the CCSN and RT templates. The peak absolute
$B$-band magnitude
$\MB$ of the SN~Ia templates is derived with equation (4) in \citet{bar12}.
The host galaxy extinction for the CCSN and RT templates is characterized with the
color excess $\Ebvh$ with a range of $0.0$ to $1.0$, and the extinction
curve of the host galaxy is assumed to be the same as in our Galaxy
\citep{pei92}. While we adopt the peak magnitudes of the observed RTs
with assuming a variation of $\pm1$~mag, the average and variation of
other parameters are taken from \citet{bar12} for SNe~Ia and \citet{dah12}
for CCSNe. The adopted templates and the range of variation are
summarized in Table~\ref{tab:template}.

Template light curves are derived with interpolating or
extrapolating the spectral energy distribution for various redshift from
$z=0$ to $z=4$ for SNe~IIn or from $z=0$ to $z=2$ for other transients with
$\Delta z=0.05$ as done in \citet{oku14,nii14frb}. We force the
template to have the information of the $g$-, $r$-, and $i$-bands at
least in an epoch and to have the information of at least 1 band in more
than 1 epoch. The explosion date $\texp$ of templates since the
occurrence time of \FRB\ are varied with $\Delta\texp=1$~day at $\texp\leq11$~day in the observer
frame. With the derived template light curves, we calculate fluxes on
Days~8, 11, and 14 and difference fluxes between Days 8 and 14 and
between Days~11 and 14.

We expedientially judge the agreement of light curves of candidates and
templates with the use of the $Q$ value of chi square
statistics as a guide. The chi square is defined as
\begin{eqnarray}
 \nonumber \chi^2 &=& \sum_{\rm difference}
  {\left(f_{\rm d,obs} - f_{\rm d,temp} \right)^2\over{\sigma_{\rm d,obs}^2}} \\
\nonumber && \times \left\{1 - \Theta \left(3 \sigma_{\rm d,obs} - \left|f_{\rm d,obs}\right|\right)
     \Theta \left(3 \sigma_{\rm d,obs} - \left|f_{\rm
					d,temp}\right|\right)\right\}\\
 &&+\sum_{\rm stacked}
  {\left(f_{\rm s,obs} - f_{\rm s,temp} \right)^2\over{\sigma_{\rm s,obs}^2}}
  \Theta \left(f_{\rm s,temp} - f_{\rm s,obs}\right),
  \label{eq:chi2}
\end{eqnarray}
where $f_{\rm i,obs}$, $f_{\rm i,temp}$, and $\sigma_{\rm i,obs}$ are the
observed flux, the template flux, and the observed standard deviation,
respectively. The suffix i represents s for the stacked images and d for
the difference images. The Heaviside function $\Theta(x)$ is
defined to be $\Theta(x) = 1$ for $x > 0$ and $0$ otherwise.
The number of observations that are used for the estimation of $\chi^2$ is
\begin{eqnarray}
\nonumber \nobs &=& \sum_{\rm difference}
  \left\{1 - \Theta \left(3 \sigma_{\rm d,obs} - \left|f_{\rm d,obs}\right|\right)
  \Theta \left(3 \sigma_{\rm d,obs} - \left|f_{\rm d,temp}\right|\right)\right\}\\
 &&+\sum_{\rm stacked}
  \Theta \left(f_{\rm s,temp} - f_{\rm s,obs}\right).
  \label{eq:nobs}
\end{eqnarray}

A degree of freedom of the chi square statistics is the difference
between the numbers of independent observations and independent
parameters. However, an identification of the independent parameters
requires great care and our aim is only to constrain possible templates.
Thus we conservatively set the degree of freedom $\nu$
as $(\nobs-1)$ because $\texp$ is obviously independent on the
other parameters related to the properties of transients.

\begin{figure}
 \begin{center}
  \includegraphics[width=8cm]{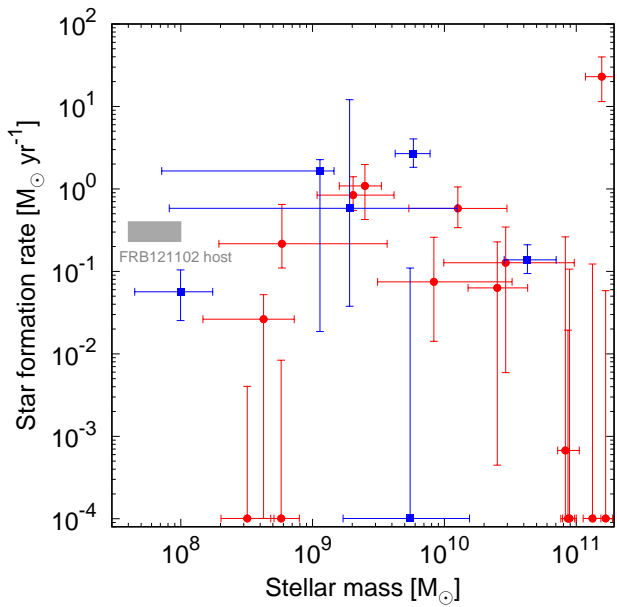} 
 \end{center}
 \caption{Stellar masses and star formation rates of optical
 counterparts of radio sources detected with VLA and GMRT (points) and
 of the host galaxy of FRB~121102 (shaded region, \cite{kok17,bas17,ten17}). The objects
 with $W1-W2\geq+0.8$~mag and $W1-W2\leq+0.8$~mag are shown in blue squares and
 red circles, respectively. The error bars represent the $68\%$
 probability.
 }
 \label{fig:radiosource}
\end{figure}

\section{Persistent optical counterparts of radio sources}
\label{sec:radio}

The radio follow-up observations of \FRB\ with VLA and GMRT detect 32
independent radio sources \citep{bha18}. Among them, 22 radio sources
have extended optical counterparts within the FWHM of the synthesis
beams in the $gri$ images obtained on Day 10 which have the best
seeing. Here, we adopt the position measured with GMRT if a source is
detected with both of VLA and GMRT because GMRT has a higher
spatial resolution than VLA. For these optical counterparts, the stellar
masses and the SFR are derived using the photometric redshift code and
summarized in Figure~\ref{fig:radiosource}. Also, we compare the
locations of radio sources having optical counterparts with the WISE
catalog and identify the sources being likely to host a AGN. There are
no radio source hosted in a galaxy identical with the host galaxy of
FRB~121102 in terms of the stellar mass and SFR, and some of the radio
sources could be emissions from AGNs according to the WISE color of the
host galaxies. Further observations are required to conclude whether the
radio sources are associated with \FRB\ or not.


\bibliographystyle{myaasjournal} 
\bibliography{ms}

\begin{thebibliography}{}
\expandafter\ifx\csname natexlab\endcsname\relax\def\natexlab#1{#1}\fi
\providecommand{\url}[1]{\href{#1}{#1}}

\bibitem[{{Akiyama} \& {Johnson}(2016)}]{aki16}
{Akiyama}, K., \& {Johnson}, M.~D. 2016, \apjl, 824, L3

\bibitem[{{Alard} \& {Lupton}(1998)}]{ala98}
{Alard}, C., \& {Lupton}, R.~H. 1998, \apj, 503, 325

\bibitem[{{Ananthakrishnan}(1995)}]{GMRT}
{Ananthakrishnan}, S. 1995, Journal of Astrophysics and Astronomy Supplement,
  16, 427

\bibitem[{{Axelrod} {et~al.}(2010){Axelrod}, {Kantor}, {Lupton}, \&
  {Pierfederici}}]{axe10}
{Axelrod}, T., {Kantor}, J., {Lupton}, R.~H., \& {Pierfederici}, F. 2010, in
  Society of Photo-Optical Instrumentation Engineers (SPIE) Conference Series,
  Vol. 7740, Society of Photo-Optical Instrumentation Engineers (SPIE)
  Conference Series, 15

\bibitem[{{Bannister} \& {Madsen}(2014)}]{ban14}
{Bannister}, K.~W., \& {Madsen}, G.~J. 2014, \mnras, 440, 353

\bibitem[{{Bannister} {et~al.}(2017){Bannister}, {Shannon}, {Macquart},
  {Flynn}, {Edwards}, {O'Neill}, {Os{\l}owski}, {Bailes}, {Zackay}, {Clarke},
  {D'Addario}, {Dodson}, {Hall}, {Jameson}, {Jones}, {Navarro}, {Trinh},
  {Allison}, {Anderson}, {Bell}, {Chippendale}, {Collier}, {Heald}, {Heywood},
  {Hotan}, {Lee-Waddell}, {Madrid}, {Marvil}, {McConnell}, {Popping},
  {Voronkov}, {Whiting}, {Allen}, {Bock}, {Brodrick}, {Cooray}, {DeBoer},
  {Diamond}, {Ekers}, {Gough}, {Hampson}, {Harvey-Smith}, {Hay}, {Hayman},
  {Jackson}, {Johnston}, {Koribalski}, {McClure-Griffiths}, {Mirtschin}, {Ng},
  {Norris}, {Pearce}, {Phillips}, {Roxby}, {Troup}, \& {Westmeier}}]{ban17}
{Bannister}, K.~W., {et~al.} 2017, \apjl, 841, L12

\bibitem[{{Barbary} {et~al.}(2012){Barbary}, {Aldering}, {Amanullah},
  {Brodwin}, {Connolly}, {Dawson}, {Doi}, {Eisenhardt}, {Faccioli}, {Fadeyev},
  {Fakhouri}, {Fruchter}, {Gilbank}, {Gladders}, {Goldhaber}, {Goobar},
  {Hattori}, {Hsiao}, {Huang}, {Ihara}, {Kashikawa}, {Koester}, {Konishi},
  {Kowalski}, {Lidman}, {Lubin}, {Meyers}, {Morokuma}, {Oda}, {Panagia},
  {Perlmutter}, {Postman}, {Ripoche}, {Rosati}, {Rubin}, {Schlegel},
  {Spadafora}, {Stanford}, {Strovink}, {Suzuki}, {Takanashi}, {Tokita},
  {Yasuda}, \& {Supernova Cosmology Project}}]{bar12}
{Barbary}, K., {et~al.} 2012, \apj, 745, 31

\bibitem[{{Bassa} {et~al.}(2016){Bassa}, {Beswick}, {Tingay}, {Keane},
  {Bhandari}, {Johnston}, {Totani}, {Tominaga}, {Yasuda}, {Stappers}, {Barr},
  {Kramer}, \& {Possenti}}]{bas16}
{Bassa}, C.~G., {et~al.} 2016, \mnras, 463, L36

\bibitem[{{Bassa} {et~al.}(2017){Bassa}, {Tendulkar}, {Adams}, {Maddox},
  {Bogdanov}, {Bower}, {Burke-Spolaor}, {Butler}, {Chatterjee}, {Cordes},
  {Hessels}, {Kaspi}, {Law}, {Marcote}, {Paragi}, {Ransom}, {Scholz},
  {Spitler}, \& {van Langevelde}}]{bas17}
---. 2017, \apjl, 843, L8

\bibitem[{{Bhandari} {et~al.}(2018){Bhandari}, {Keane}, {Barr}, {Jameson},
  {Petroff}, {Johnston}, {Bailes}, {Bhat}, {Burgay}, {Burke-Spolaor}, {Caleb},
  {Eatough}, {Flynn}, {Green}, {Jankowski}, {Kramer}, {Krishnan}, {Morello},
  {Possenti}, {Stappers}, {Tiburzi}, {van Straten}, {Andreoni}, {Butterley},
  {Chandra}, {Cooke}, {Corongiu}, {Coward}, {Dhillon}, {Dodson}, {Hardy},
  {Howell}, {Jaroenjittichai}, {Klotz}, {Littlefair}, {Marsh}, {Mickaliger},
  {Muxlow}, {Perrodin}, {Pritchard}, {Sawangwit}, {Terai}, {Tominaga}, {Torne},
  {Totani}, {Trois}, {Turpin}, {Niino}, {Wilson}, {Albert}, {Andr{\'e}},
  {Anghinolfi}, {Anton}, {Ardid}, {Aubert}, {Avgitas}, {Baret},
  {Barrios-Mart{\'{\i}}}, {Basa}, {Belhorma}, {Bertin}, {Biagi}, {Bormuth},
  {Bourret}, {Bouwhuis}, {Br{\^a}nza{\c s}}, {Bruijn}, {Brunner}, {Busto},
  {Capone}, {Caramete}, {Carr}, {Celli}, {Moursli}, {Chiarusi}, {Circella},
  {Coelho}, {Coleiro}, {Coniglione}, {Costantini}, {Coyle}, {Creusot},
  {D{\'{\i}}az}, {Deschamps}, {De Bonis}, {Distefano}, {Palma}, {Domi},
  {Donzaud}, {Dornic}, {Drouhin}, {Eberl}, {Bojaddaini}, {Khayati},
  {Els{\"a}sser}, {Enzenh{\"o}fer}, {Ettahiri}, {Fassi}, {Felis}, {Fusco},
  {Gay}, {Giordano}, {Glotin}, {Gregoire}, {Gracia-Ruiz}, {Graf}, {Hallmann},
  {van Haren}, {Heijboer}, {Hello}, {Hern{\'a}ndez-Rey}, {H{\"o}{\ss}l},
  {Hofest{\"a}dt}, {Hugon}, {Illuminati}, {James}, {de Jong}, {Jongen},
  {Kadler}, {Kalekin}, {Katz}, {Kie{\ss}ling}, {Kouchner}, {Kreter},
  {Kreykenbohm}, {Kulikovskiy}, {Lachaud}, {Lahmann}, {Lef{\`e}vre}, {Leonora},
  {Loucatos}, {Marcelin}, {Margiotta}, {Marinelli}, {Mart{\'{\i}}nez-Mora},
  {Mele}, {Melis}, {Michael}, {Migliozzi}, {Moussa}, {Navas}, {Nezri},
  {Organokov}, {P{\v a}v{\v a}la{\c s}}, {Pellegrino}, {Perrina}, {Piattelli},
  {Popa}, {Pradier}, {Quinn}, {Racca}, {Riccobene}, {S{\'a}nchez-Losa},
  {Salda{\~n}a}, {Salvadori}, {Samtleben}, {Sanguineti}, {Sapienza},
  {Sch{\"u}ssler}, {Sieger}, {Spurio}, {Stolarczyk}, {Taiuti}, {Tayalati},
  {Trovato}, {Turpin}, {T{\"o}nnis}, {Vallage}, {Van Elewyck}, {Versari},
  {Vivolo}, {Vizzocca}, {Wilms}, {Zornoza}, \& {Z{\'u}{\~n}iga}}]{bha18}
{Bhandari}, S., {et~al.} 2018, \mnras, 475, 1427

\bibitem[{{Boller} {et~al.}(2016){Boller}, {Freyberg}, {Tr{\"u}mper}, {Haberl},
  {Voges}, \& {Nandra}}]{ROSATcatalog}
{Boller}, T., {Freyberg}, M.~J., {Tr{\"u}mper}, J., {Haberl}, F., {Voges}, W.,
  \& {Nandra}, K. 2016, \aap, 588, A103

\bibitem[{{Bosch} {et~al.}(2017){Bosch}, {Armstrong}, {Bickerton}, {Furusawa},
  {Ikeda}, {Koike}, {Lupton}, {Mineo}, {Price}, {Takata}, {Tanaka}, {Yasuda},
  {AlSayyad}, {Becker}, {Coulton}, {Coupon}, {Garmilla}, {Huang}, {Krughoff},
  {Lang}, {Leauthaud}, {Lim}, {Lust}, {MacArthur}, {Mandelbaum}, {Miyatake},
  {Miyazaki}, {Murata}, {More}, {Okura}, {Owen}, {Swinbank}, {Strauss},
  {Yamada}, \& {Yamanoi}}]{bos17}
{Bosch}, J., {et~al.} 2017, ArXiv e-prints, arXiv:1705.06766

\bibitem[{{Brooks} {et~al.}(2017){Brooks}, {Schwab}, {Bildsten}, {Quataert},
  {Paxton}, {Blinnikov}, \& {Sorokina}}]{bro17}
{Brooks}, J., {Schwab}, J., {Bildsten}, L., {Quataert}, E., {Paxton}, B.,
  {Blinnikov}, S., \& {Sorokina}, E. 2017, \apj, 850, 127

\bibitem[{{Caleb} {et~al.}(2017){Caleb}, {Flynn}, {Bailes}, {Barr}, {Bateman},
  {Bhandari}, {Campbell-Wilson}, {Farah}, {Green}, {Hunstead}, {Jameson},
  {Jankowski}, {Keane}, {Parthasarathy}, {Ravi}, {Rosado}, {van Straten}, \&
  {Venkatraman Krishnan}}]{cal17}
{Caleb}, M., {et~al.} 2017, \mnras, 468, 3746

\bibitem[{{Chambers} {et~al.}(2016){Chambers}, {Magnier}, {Metcalfe},
  {Flewelling}, {Huber}, {Waters}, {Denneau}, {Draper}, {Farrow}, {Finkbeiner},
  {Holmberg}, {Koppenhoefer}, {Price}, {Saglia}, {Schlafly}, {Smartt},
  {Sweeney}, {Wainscoat}, {Burgett}, {Grav}, {Heasley}, {Hodapp}, {Jedicke},
  {Kaiser}, {Kudritzki}, {Luppino}, {Lupton}, {Monet}, {Morgan}, {Onaka},
  {Stubbs}, {Tonry}, {Banados}, {Bell}, {Bender}, {Bernard}, {Botticella},
  {Casertano}, {Chastel}, {Chen}, {Chen}, {Cole}, {Deacon}, {Frenk},
  {Fitzsimmons}, {Gezari}, {Goessl}, {Goggia}, {Goldman}, {Grebel}, {Hambly},
  {Hasinger}, {Heavens}, {Heckman}, {Henderson}, {Henning}, {Holman}, {Hopp},
  {Ip}, {Isani}, {Keyes}, {Koekemoer}, {Kotak}, {Long}, {Lucey}, {Liu},
  {Martin}, {McLean}, {Morganson}, {Murphy}, {Nieto-Santisteban}, {Norberg},
  {Peacock}, {Pier}, {Postman}, {Primak}, {Rae}, {Rest}, {Riess}, {Riffeser},
  {Rix}, {Roser}, {Schilbach}, {Schultz}, {Scolnic}, {Szalay}, {Seitz},
  {Shiao}, {Small}, {Smith}, {Soderblom}, {Taylor}, {Thakar}, {Thiel},
  {Thilker}, {Urata}, {Valenti}, {Walter}, {Watters}, {Werner}, {White},
  {Wood-Vasey}, \& {Wyse}}]{ps1survey}
{Chambers}, K.~C., {et~al.} 2016, ArXiv e-prints, arXiv:1612.05560

\bibitem[{{Champion} {et~al.}(2016){Champion}, {Petroff}, {Kramer}, {Keith},
  {Bailes}, {Barr}, {Bates}, {Bhat}, {Burgay}, {Burke-Spolaor}, {Flynn},
  {Jameson}, {Johnston}, {Ng}, {Levin}, {Possenti}, {Stappers}, {van Straten},
  {Thornton}, {Tiburzi}, \& {Lyne}}]{cha16}
{Champion}, D.~J., {et~al.} 2016, \mnras, 460, L30

\bibitem[{{Chatterjee} {et~al.}(2017){Chatterjee}, {Law}, {Wharton},
  {Burke-Spolaor}, {Hessels}, {Bower}, {Cordes}, {Tendulkar}, {Bassa},
  {Demorest}, {Butler}, {Seymour}, {Scholz}, {Abruzzo}, {Bogdanov}, {Kaspi},
  {Keimpema}, {Lazio}, {Marcote}, {McLaughlin}, {Paragi}, {Ransom}, {Rupen},
  {Spitler}, \& {van Langevelde}}]{cha17}
{Chatterjee}, S., {et~al.} 2017, \nat, 541, 58

\bibitem[{{Condon} {et~al.}(1998){Condon}, {Cotton}, {Greisen}, {Yin},
  {Perley}, {Taylor}, \& {Broderick}}]{NVSScatalog}
{Condon}, J.~J., {Cotton}, W.~D., {Greisen}, E.~W., {Yin}, Q.~F., {Perley},
  R.~A., {Taylor}, G.~B., \& {Broderick}, J.~J. 1998, \aj, 115, 1693

\bibitem[{{Connor} {et~al.}(2016){Connor}, {Sievers}, \& {Pen}}]{con16}
{Connor}, L., {Sievers}, J., \& {Pen}, U.-L. 2016, \mnras, 458, L19

\bibitem[{{Cordes} \& {Wasserman}(2016)}]{cor16}
{Cordes}, J.~M., \& {Wasserman}, I. 2016, \mnras, 457, 232

\bibitem[{{Dahlen} {et~al.}(2012){Dahlen}, {Strolger}, {Riess}, {Mattila},
  {Kankare}, \& {Mobasher}}]{dah12}
{Dahlen}, T., {Strolger}, L.-G., {Riess}, A.~G., {Mattila}, S., {Kankare}, E.,
  \& {Mobasher}, B. 2012, \apj, 757, 70

\bibitem[{{Drout} {et~al.}(2014){Drout}, {Chornock}, {Soderberg}, {Sanders},
  {McKinnon}, {Rest}, {Foley}, {Milisavljevic}, {Margutti}, {Berger},
  {Calkins}, {Fong}, {Gezari}, {Huber}, {Kankare}, {Kirshner}, {Leibler},
  {Lunnan}, {Mattila}, {Marion}, {Narayan}, {Riess}, {Roth}, {Scolnic},
  {Smartt}, {Tonry}, {Burgett}, {Chambers}, {Hodapp}, {Jedicke}, {Kaiser},
  {Magnier}, {Metcalfe}, {Morgan}, {Price}, \& {Waters}}]{dro14}
{Drout}, M.~R., {et~al.} 2014, \apj, 794, 23

\bibitem[{{Egorov} \& {Postnov}(2009)}]{ego09}
{Egorov}, A.~E., \& {Postnov}, K.~A. 2009, Astronomy Letters, 35, 241

\bibitem[{{Falcke} \& {Rezzolla}(2014)}]{fal14}
{Falcke}, H., \& {Rezzolla}, L. 2014, \aap, 562, A137

\bibitem[{{Hsiao} {et~al.}(2007){Hsiao}, {Conley}, {Howell}, {Sullivan},
  {Pritchet}, {Carlberg}, {Nugent}, \& {Phillips}}]{hsi07}
{Hsiao}, E.~Y., {Conley}, A., {Howell}, D.~A., {Sullivan}, M., {Pritchet},
  C.~J., {Carlberg}, R.~G., {Nugent}, P.~E., \& {Phillips}, M.~M. 2007, \apj,
  663, 1187

\bibitem[{{Ivezic} {et~al.}(2008){Ivezic}, {Tyson}, {Allsman}, {Andrew},
  {Angel}, \& {for the LSST Collaboration}}]{ive08}
{Ivezic}, Z., {Tyson}, J.~A., {Allsman}, R., {Andrew}, J., {Angel}, R., \& {for
  the LSST Collaboration}. 2008, ArXiv e-prints, arXiv:0805.2366

\bibitem[{{Johnston} {et~al.}(2017){Johnston}, {Keane}, {Bhandari}, {Macquart},
  {Tingay}, {Barr}, {Bassa}, {Beswick}, {Burgay}, {Chandra}, {Honma}, {Kramer},
  {Petroff}, {Possenti}, {Stappers}, \& {Sugai}}]{joh17}
{Johnston}, S., {et~al.} 2017, \mnras, 465, 2143

\bibitem[{{Juri{\'c}} {et~al.}(2015){Juri{\'c}}, {Kantor}, {Lim}, {Lupton},
  {Dubois-Felsmann}, {Jenness}, {Axelrod}, {Aleksi{\'c}}, {Allsman},
  {AlSayyad}, {Alt}, {Armstrong}, {Basney}, {Becker}, {Becla}, {Bickerton},
  {Biswas}, {Bosch}, {Boutigny}, {Carrasco Kind}, {Ciardi}, {Connolly},
  {Daniel}, {Daues}, {Economou}, {Chiang}, {Fausti}, {Fisher-Levine},
  {Freemon}, {Gee}, {Gris}, {Hernandez}, {Hoblitt}, {Ivezi{\'c}}, {Jammes},
  {Jevremovi{\'c}}, {Jones}, {Bryce Kalmbach}, {Kasliwal}, {Krughoff}, {Lang},
  {Lurie}, {Lust}, {Mullally}, {MacArthur}, {Melchior}, {Moeyens}, {Nidever},
  {Owen}, {Parejko}, {Peterson}, {Petravick}, {Pietrowicz}, {Price}, {Reiss},
  {Shaw}, {Sick}, {Slater}, {Strauss}, {Sullivan}, {Swinbank}, {Van Dyk},
  {Vuj{\v c}i{\'c}}, {Withers}, {Yoachim}, \& {LSST Project}}]{jur15}
{Juri{\'c}}, M., {et~al.} 2015, ArXiv e-prints, arXiv:1512.07914

\bibitem[{{Kashiyama} {et~al.}(2013){Kashiyama}, {Ioka}, \&
  {M{\'e}sz{\'a}ros}}]{kas13FRB}
{Kashiyama}, K., {Ioka}, K., \& {M{\'e}sz{\'a}ros}, P. 2013, \apjl, 776, L39

\bibitem[{{Kashiyama} \& {Murase}(2017)}]{kas17}
{Kashiyama}, K., \& {Murase}, K. 2017, \apjl, 839, L3

\bibitem[{{Keane} \& {Petroff}(2015)}]{kea15}
{Keane}, E.~F., \& {Petroff}, E. 2015, \mnras, 447, 2852

\bibitem[{{Keane} {et~al.}(2012){Keane}, {Stappers}, {Kramer}, \&
  {Lyne}}]{kea12}
{Keane}, E.~F., {Stappers}, B.~W., {Kramer}, M., \& {Lyne}, A.~G. 2012, \mnras,
  425, L71

\bibitem[{{Keane} {et~al.}(2016){Keane}, {Johnston}, {Bhandari}, {Barr},
  {Bhat}, {Burgay}, {Caleb}, {Flynn}, {Jameson}, {Kramer}, {Petroff},
  {Possenti}, {van Straten}, {Bailes}, {Burke-Spolaor}, {Eatough}, {Stappers},
  {Totani}, {Honma}, {Furusawa}, {Hattori}, {Morokuma}, {Niino}, {Sugai},
  {Terai}, {Tominaga}, {Yamasaki}, {Yasuda}, {Allen}, {Cooke}, {Jencson},
  {Kasliwal}, {Kaplan}, {Tingay}, {Williams}, {Wayth}, {Chandra}, {Perrodin},
  {Berezina}, {Mickaliger}, \& {Bassa}}]{kea16}
{Keane}, E.~F., {et~al.} 2016, \nat, 530, 453

\bibitem[{{Keane} {et~al.}(2018){Keane}, {Barr}, {Jameson}, {Morello}, {Caleb},
  {Bhandari}, {Petroff}, {Possenti}, {Burgay}, {Tiburzi}, {Bailes}, {Bhat},
  {Burke-Spolaor}, {Eatough}, {Flynn}, {Jankowski}, {Johnston}, {Kramer},
  {Levin}, {Ng}, {van Straten}, \& {Krishnan}}]{kea17}
---. 2018, \mnras, 473, 116

\bibitem[{{Kokubo} {et~al.}(2017){Kokubo}, {Mitsuda}, {Sugai}, {Ozaki},
  {Minowa}, {Hattori}, {Hayano}, {Matsubayashi}, {Shimono}, {Sako}, \&
  {Doi}}]{kok17}
{Kokubo}, M., {et~al.} 2017, \apj, 844, 95

\bibitem[{{Lorimer} {et~al.}(2007){Lorimer}, {Bailes}, {McLaughlin},
  {Narkevic}, \& {Crawford}}]{lor07}
{Lorimer}, D.~R., {Bailes}, M., {McLaughlin}, M.~A., {Narkevic}, D.~J., \&
  {Crawford}, F. 2007, Science, 318, 777

\bibitem[{{Macquart} \& {Ekers}(2018)}]{mac18}
{Macquart}, J.-P., \& {Ekers}, R.~D. 2018, \mnras, 474, 1900

\bibitem[{{Mainzer} {et~al.}(2011){Mainzer}, {Bauer}, {Grav}, {Masiero},
  {Cutri}, {Dailey}, {Eisenhardt}, {McMillan}, {Wright}, {Walker}, {Jedicke},
  {Spahr}, {Tholen}, {Alles}, {Beck}, {Brandenburg}, {Conrow}, {Evans},
  {Fowler}, {Jarrett}, {Marsh}, {Masci}, {McCallon}, {Wheelock}, {Wittman},
  {Wyatt}, {DeBaun}, {Elliott}, {Elsbury}, {Gautier}, {Gomillion}, {Leisawitz},
  {Maleszewski}, {Micheli}, \& {Wilkins}}]{NEOWISEcatalog}
{Mainzer}, A., {et~al.} 2011, \apj, 731, 53

\bibitem[{{Manchester} {et~al.}(2001){Manchester}, {Lyne}, {Camilo}, {Bell},
  {Kaspi}, {D'Amico}, {McKay}, {Crawford}, {Stairs}, {Possenti}, {Kramer}, \&
  {Sheppard}}]{man01}
{Manchester}, R.~N., {et~al.} 2001, \mnras, 328, 17

\bibitem[{{Marcote} {et~al.}(2017){Marcote}, {Paragi}, {Hessels}, {Keimpema},
  {van Langevelde}, {Huang}, {Bassa}, {Bogdanov}, {Bower}, {Burke-Spolaor},
  {Butler}, {Campbell}, {Chatterjee}, {Cordes}, {Demorest}, {Garrett}, {Ghosh},
  {Kaspi}, {Law}, {Lazio}, {McLaughlin}, {Ransom}, {Salter}, {Scholz},
  {Seymour}, {Siemion}, {Spitler}, {Tendulkar}, \& {Wharton}}]{mar17}
{Marcote}, B., {et~al.} 2017, \apjl, 834, L8

\bibitem[{{Masui} {et~al.}(2015){Masui}, {Lin}, {Sievers}, {Anderson}, {Chang},
  {Chen}, {Ganguly}, {Jarvis}, {Kuo}, {Li}, {Liao}, {McLaughlin}, {Pen},
  {Peterson}, {Roman}, {Timbie}, {Voytek}, \& {Yadav}}]{mas15}
{Masui}, K., {et~al.} 2015, \nat, 528, 523

\bibitem[{{McQuinn}(2014)}]{mcq14}
{McQuinn}, M. 2014, \apjl, 780, L33

\bibitem[{{Miyazaki} {et~al.}(2012){Miyazaki}, {Komiyama}, {Nakaya}, {Kamata},
  {Doi}, {Hamana}, {Karoji}, {Furusawa}, {Kawanomoto}, {Morokuma}, {Ishizuka},
  {Nariai}, {Tanaka}, {Uraguchi}, {Utsumi}, {Obuchi}, {Okura}, {Oguri},
  {Takata}, {Tomono}, {Kurakami}, {Namikawa}, {Usuda}, {Yamanoi}, {Terai},
  {Uekiyo}, {Yamada}, {Koike}, {Aihara}, {Fujimori}, {Mineo}, {Miyatake},
  {Yasuda}, {Nishizawa}, {Saito}, {Tanaka}, {Uchida}, {Katayama}, {Wang},
  {Chen}, {Lupton}, {Loomis}, {Bickerton}, {Price}, {Gunn}, {Suzuki},
  {Miyazaki}, {Muramatsu}, {Yamamoto}, {Endo}, {Ezaki}, {Itoh}, {Miwa},
  {Yokota}, {Matsuda}, {Ebinuma}, \& {Takeshi}}]{miy12}
{Miyazaki}, S., {et~al.} 2012, in Society of Photo-Optical Instrumentation
  Engineers (SPIE) Conference Series, Vol. 8446, Society of Photo-Optical
  Instrumentation Engineers (SPIE) Conference Series, 0

\bibitem[{{Miyazaki} {et~al.}(2018){Miyazaki}, {Komiyama}, {Kawanomoto}, {Doi},
  {Furusawa}, {Hamana}, {Hayashi}, {Ikeda}, {Kamata}, {Karoji}, {Koike},
  {Kurakami}, {Miyama}, {Morokuma}, {Nakata}, {Namikawa}, {Nakaya}, {Nariai},
  {Obuchi}, {Oishi}, {Okada}, {Okura}, {Tait}, {Takata}, {Tanaka}, {Tanaka},
  {Terai}, {Tomono}, {Uraguchi}, {Usuda}, {Utsumi}, {Yamada}, {Yamanoi},
  {Aihara}, {Fujimori}, {Mineo}, {Miyatake}, {Oguri}, {Uchida}, {Tanaka},
  {Yasuda}, {Takada}, {Murayama}, {Nishizawa}, {Sugiyama}, {Chiba}, {Futamase},
  {Wang}, {Chen}, {Ho}, {Liaw}, {Chiu}, {Ho}, {Lai}, {Lee}, {Jeng}, {Iwamura},
  {Armstrong}, {Bickerton}, {Bosch}, {Gunn}, {Lupton}, {Loomis}, {Price},
  {Smith}, {Strauss}, {Turner}, {Suzuki}, {Miyazaki}, {Muramatsu}, {Yamamoto},
  {Endo}, {Ezaki}, {Ito}, {Kawaguchi}, {Sofuku}, {Taniike}, {Akutsu}, {Dojo},
  {Kasumi}, {Matsuda}, {Imoto}, {Miwa}, {Suzuki}, {Takeshi}, \&
  {Yokota}}]{miy18hsc}
{Miyazaki}, S., {et~al.} 2018, \pasj, 70, S1

\bibitem[{{Morii} {et~al.}(2016){Morii}, {Ikeda}, {Tominaga}, {Tanaka},
  {Morokuma}, {Ishiguro}, {Yamato}, {Ueda}, {Suzuki}, {Yasuda}, \&
  {Yoshida}}]{morii16}
{Morii}, M., {et~al.} 2016, \pasj, 68, 104

\bibitem[{{Moriya}(2016)}]{moriya16}
{Moriya}, T.~J. 2016, \apjl, 830, L38

\bibitem[{{Moriya} {et~al.}(2017){Moriya}, {Mazzali}, {Tominaga}, {Hachinger},
  {Blinnikov}, {Tauris}, {Takahashi}, {Tanaka}, {Langer}, \&
  {Podsiadlowski}}]{moriya17}
{Moriya}, T.~J., {et~al.} 2017, \mnras, 466, 2085

\bibitem[{{Morokuma} {et~al.}(2008){Morokuma}, {Doi}, {Yasuda}, {Akiyama},
  {Sekiguchi}, {Furusawa}, {Ueda}, {Totani}, {Oda}, {Nagao}, {Kashikawa},
  {Murayama}, {Ouchi}, {Watson}, {Richmond}, {Lidman}, {Perlmutter},
  {Spadafora}, {Aldering}, {Wang}, {Hook}, \& {Knop}}]{mor08variable}
{Morokuma}, T., {et~al.} 2008, \apj, 676, 163

\bibitem[{{Niino} {et~al.}(2014){Niino}, {Totani}, \& {Okumura}}]{nii14frb}
{Niino}, Y., {Totani}, T., \& {Okumura}, J.~E. 2014, \pasj, 66, L9

\bibitem[{{Nugent} {et~al.}(2002){Nugent}, {Kim}, \& {Perlmutter}}]{nug02}
{Nugent}, P., {Kim}, A., \& {Perlmutter}, S. 2002, \pasp, 114, 803

\bibitem[{{Okumura} {et~al.}(2014){Okumura}, {Ihara}, {Doi}, {Morokuma},
  {Pain}, {Totani}, {Barbary}, {Takanashi}, {Yasuda}, {Aldering}, {Dawson},
  {Goldhaber}, {Hook}, {Lidman}, {Perlmutter}, {Spadafora}, {Suzuki}, \&
  {Wang}}]{oku14}
{Okumura}, J.~E., {et~al.} 2014, \pasj, 66, 49

\bibitem[{{Pei}(1992)}]{pei92}
{Pei}, Y.~C. 1992, \apj, 395, 130

\bibitem[{{Pen} \& {Connor}(2015)}]{pen15}
{Pen}, U.-L., \& {Connor}, L. 2015, \apj, 807, 179

\bibitem[{{Petroff} {et~al.}(2015{\natexlab{a}}){Petroff}, {Johnston}, {Keane},
  {van Straten}, {Bailes}, {Barr}, {Barsdell}, {Burke-Spolaor}, {Caleb},
  {Champion}, {Flynn}, {Jameson}, {Kramer}, {Ng}, {Possenti}, \&
  {Stappers}}]{pet15}
{Petroff}, E., {et~al.} 2015{\natexlab{a}}, \mnras, 454, 457

\bibitem[{{Petroff} {et~al.}(2015{\natexlab{b}}){Petroff}, {Bailes}, {Barr},
  {Barsdell}, {Bhat}, {Bian}, {Burke-Spolaor}, {Caleb}, {Champion}, {Chandra},
  {Da Costa}, {Delvaux}, {Flynn}, {Gehrels}, {Greiner}, {Jameson}, {Johnston},
  {Kasliwal}, {Keane}, {Keller}, {Kocz}, {Kramer}, {Leloudas}, {Malesani},
  {Mulchaey}, {Ng}, {Ofek}, {Perley}, {Possenti}, {Schmidt}, {Shen},
  {Stappers}, {Tisserand}, {van Straten}, \& {Wolf}}]{pet15frb140514}
---. 2015{\natexlab{b}}, \mnras, 447, 246

\bibitem[{{Petroff} {et~al.}(2016){Petroff}, {Barr}, {Jameson}, {Keane},
  {Bailes}, {Kramer}, {Morello}, {Tabbara}, \& {van Straten}}]{pet16}
---. 2016, \pasa, 33, e045

\bibitem[{{Pian} {et~al.}(2017){Pian}, {D{\rsquo}Avanzo}, {Benetti},
  {Branchesi}, {Brocato}, {Campana}, {Cappellaro}, {Covino}, {D{\rsquo}Elia},
  {Fynbo}, {Getman}, {Ghirlanda}, {Ghisellini}, {Grado}, {Greco}, {Hjorth},
  {Kouveliotou}, {Levan}, {Limatola}, {Malesani}, {Mazzali}, {Melandri},
  {M{\o}ller}, {Nicastro}, {Palazzi}, {Piranomonte}, {Rossi}, {Salafia},
  {Selsing}, {Stratta}, {Tanaka}, {Tanvir}, {Tomasella}, {Watson}, {Yang},
  {Amati}, {Antonelli}, {Ascenzi}, {Bernardini}, {Bo{\"e}r}, {Bufano},
  {Bulgarelli}, {Capaccioli}, {Casella}, {Castro-Tirado}, {Chassande-Mottin},
  {Ciolfi}, {Copperwheat}, {Dadina}, {De Cesare}, {di Paola}, {Fan}, {Gendre},
  {Giuffrida}, {Giunta}, {Hunt}, {Israel}, {Jin}, {Kasliwal}, {Klose}, {Lisi},
  {Longo}, {Maiorano}, {Mapelli}, {Masetti}, {Nava}, {Patricelli}, {Perley},
  {Pescalli}, {Piran}, {Possenti}, {Pulone}, {Razzano}, {Salvaterra},
  {Schipani}, {Spera}, {Stamerra}, {Stella}, {Tagliaferri}, {Testa}, {Troja},
  {Turatto}, {Vergani}, \& {Vergani}}]{pian17}
{Pian}, E., {et~al.} 2017, \nat, 551, 67

\bibitem[{{Piro}(2016)}]{pir16}
{Piro}, A.~L. 2016, \apjl, 824, L32

\bibitem[{{Planck Collaboration} {et~al.}(2016){Planck Collaboration}, {Ade},
  {Aghanim}, {Aller}, {Aller}, {Arnaud}, {Aumont}, {Baccigalupi}, {Banday},
  {Barreiro}, {Bartolo}, {Battaner}, {Benabed}, {Benoit-L{\'e}vy}, {Bernard},
  {Bersanelli}, {Bielewicz}, {Bonaldi}, {Bonavera}, {Bond}, {Borrill},
  {Bouchet}, {Burigana}, {Calabrese}, {Catalano}, {Chiang}, {Christensen},
  {Clements}, {Colombo}, {Couchot}, {Crill}, {Curto}, {Cuttaia}, {Danese},
  {Davies}, {Davis}, {de Bernardis}, {de Rosa}, {de Zotti}, {Delabrouille},
  {Dickinson}, {Diego}, {Dole}, {Donzelli}, {Dor{\'e}}, {Ducout}, {Dupac},
  {Efstathiou}, {Elsner}, {Eriksen}, {Finelli}, {Forni}, {Frailis}, {Fraisse},
  {Franceschi}, {Galeotta}, {Galli}, {Ganga}, {Giard}, {Giraud-H{\'e}raud},
  {Gjerl{\o}w}, {Gonz{\'a}lez-Nuevo}, {G{\'o}rski}, {Gruppuso}, {Gurwell},
  {Hansen}, {Harrison}, {Henrot-Versill{\'e}}, {Hern{\'a}ndez-Monteagudo},
  {Hildebrandt}, {Hobson}, {Hornstrup}, {Hovatta}, {Hovest}, {Huffenberger},
  {Hurier}, {Jaffe}, {Jaffe}, {J{\"a}rvel{\"a}}, {Keih{\"a}nen}, {Keskitalo},
  {Kisner}, {Kneissl}, {Knoche}, {Kunz}, {Kurki-Suonio}, {L{\"a}hteenm{\"a}ki},
  {Lamarre}, {Lasenby}, {Lattanzi}, {Lawrence}, {Leonardi}, {Levrier},
  {Liguori}, {Lilje}, {Linden-V{\o}rnle}, {L{\'o}pez-Caniego}, {Lubin},
  {Mac{\'{\i}}as-P{\'e}rez}, {Maffei}, {Maino}, {Mandolesi}, {Maris}, {Martin},
  {Mart{\'{\i}}nez-Gonz{\'a}lez}, {Masi}, {Matarrese}, {Max-Moerbeck},
  {Meinhold}, {Melchiorri}, {Mennella}, {Migliaccio}, {Mingaliev},
  {Miville-Desch{\^e}nes}, {Moneti}, {Montier}, {Morgante}, {Mortlock},
  {Munshi}, {Murphy}, {Nati}, {Natoli}, {Nieppola}, {Noviello}, {Novikov},
  {Novikov}, {Pagano}, {Pajot}, {Paoletti}, {Partridge}, {Pasian}, {Pearson},
  {Perdereau}, {Perotto}, {Pettorino}, {Piacentini}, {Piat}, {Pierpaoli},
  {Plaszczynski}, {Pointecouteau}, {Polenta}, {Pratt}, {Ramakrishnan},
  {Rastorgueva-Foi}, {S Readhead}, {Reinecke}, {Remazeilles}, {Renault},
  {Renzi}, {Richards}, {Ristorcelli}, {Rocha}, {Rossetti}, {Roudier},
  {Rubi{\~n}o-Mart{\'{\i}}n}, {Rusholme}, {Sandri}, {Savelainen}, {Savini},
  {Scott}, {Sotnikova}, {Stolyarov}, {Sunyaev}, {Sutton}, {Suur-Uski},
  {Sygnet}, {Tammi}, {Tauber}, {Terenzi}, {Toffolatti}, {Tomasi}, {Tornikoski},
  {Tristram}, {Tucci}, {T{\"u}rler}, {Valenziano}, {Valiviita}, {Valtaoja},
  {Van Tent}, {Vielva}, {Villa}, {Wade}, {Wehrle}, {Wehus}, {Yvon}, {Zacchei},
  \& {Zonca}}]{planck16radioagn}
{Planck Collaboration}, {et~al.} 2016, \aap, 596, A106

\bibitem[{{Ravi} {et~al.}(2015){Ravi}, {Shannon}, \& {Jameson}}]{rav15}
{Ravi}, V., {Shannon}, R.~M., \& {Jameson}, A. 2015, \apjl, 799, L5

\bibitem[{{Ravi} {et~al.}(2016){Ravi}, {Shannon}, {Bailes}, {Bannister},
  {Bhandari}, {Bhat}, {Burke-Spolaor}, {Caleb}, {Flynn}, {Jameson}, {Johnston},
  {Keane}, {Kerr}, {Tiburzi}, {Tuntsov}, \& {Vedantham}}]{rav16}
{Ravi}, V., {et~al.} 2016, Science, 354, 1249

\bibitem[{{Rosen} {et~al.}(2016){Rosen}, {Webb}, {Watson}, {Ballet}, {Barret},
  {Braito}, {Carrera}, {Ceballos}, {Coriat}, {Della Ceca}, {Denkinson},
  {Esquej}, {Farrell}, {Freyberg}, {Gris{\'e}}, {Guillout}, {Heil},
  {Koliopanos}, {Law-Green}, {Lamer}, {Lin}, {Martino}, {Michel}, {Motch},
  {Nebot Gomez-Moran}, {Page}, {Page}, {Page}, {Pakull}, {Pye}, {Read},
  {Rodriguez}, {Sakano}, {Saxton}, {Schwope}, {Scott}, {Sturm}, {Traulsen},
  {Yershov}, \& {Zolotukhin}}]{XMMcatalog}
{Rosen}, S.~R., {et~al.} 2016, \aap, 590, A1

\bibitem[{{Schlafly} \& {Finkbeiner}(2011)}]{sch11}
{Schlafly}, E.~F., \& {Finkbeiner}, D.~P. 2011, \apj, 737, 103

\bibitem[{{Spitler} {et~al.}(2014){Spitler}, {Cordes}, {Hessels}, {Lorimer},
  {McLaughlin}, {Chatterjee}, {Crawford}, {Deneva}, {Kaspi}, {Wharton},
  {Allen}, {Bogdanov}, {Brazier}, {Camilo}, {Freire}, {Jenet},
  {Karako-Argaman}, {Knispel}, {Lazarus}, {Lee}, {van Leeuwen}, {Lynch},
  {Ransom}, {Scholz}, {Siemens}, {Stairs}, {Stovall}, {Swiggum},
  {Venkataraman}, {Zhu}, {Aulbert}, \& {Fehrmann}}]{spi14}
{Spitler}, L.~G., {et~al.} 2014, \apj, 790, 101

\bibitem[{{Spitler} {et~al.}(2016){Spitler}, {Scholz}, {Hessels}, {Bogdanov},
  {Brazier}, {Camilo}, {Chatterjee}, {Cordes}, {Crawford}, {Deneva}, {Ferdman},
  {Freire}, {Kaspi}, {Lazarus}, {Lynch}, {Madsen}, {McLaughlin}, {Patel},
  {Ransom}, {Seymour}, {Stairs}, {Stappers}, {van Leeuwen}, \& {Zhu}}]{spi16}
---. 2016, \nat, 531, 202

\bibitem[{{Stern} {et~al.}(2012){Stern}, {Assef}, {Benford}, {Blain}, {Cutri},
  {Dey}, {Eisenhardt}, {Griffith}, {Jarrett}, {Lake}, {Masci}, {Petty},
  {Stanford}, {Tsai}, {Wright}, {Yan}, {Harrison}, \& {Madsen}}]{ste12}
{Stern}, D., {et~al.} 2012, \apj, 753, 30

\bibitem[{{Suwa} {et~al.}(2015){Suwa}, {Yoshida}, {Shibata}, {Umeda}, \&
  {Takahashi}}]{suw15}
{Suwa}, Y., {Yoshida}, T., {Shibata}, M., {Umeda}, H., \& {Takahashi}, K. 2015,
  \mnras, 454, 3073

\bibitem[{{Tanaka}(2015)}]{tanaka15photoz}
{Tanaka}, M. 2015, \apj, 801, 20

\bibitem[{{Tauris} {et~al.}(2015){Tauris}, {Langer}, \&
  {Podsiadlowski}}]{tau15}
{Tauris}, T.~M., {Langer}, N., \& {Podsiadlowski}, P. 2015, \mnras, 451, 2123

\bibitem[{{Tendulkar} {et~al.}(2017){Tendulkar}, {Bassa}, {Cordes}, {Bower},
  {Law}, {Chatterjee}, {Adams}, {Bogdanov}, {Burke-Spolaor}, {Butler},
  {Demorest}, {Hessels}, {Kaspi}, {Lazio}, {Maddox}, {Marcote}, {McLaughlin},
  {Paragi}, {Ransom}, {Scholz}, {Seymour}, {Spitler}, {van Langevelde}, \&
  {Wharton}}]{ten17}
{Tendulkar}, S.~P., {et~al.} 2017, \apjl, 834, L7

\bibitem[{{Thornton} {et~al.}(2013){Thornton}, {Stappers}, {Bailes},
  {Barsdell}, {Bates}, {Bhat}, {Burgay}, {Burke-Spolaor}, {Champion}, {Coster},
  {D'Amico}, {Jameson}, {Johnston}, {Keith}, {Kramer}, {Levin}, {Milia}, {Ng},
  {Possenti}, \& {van Straten}}]{tho13}
{Thornton}, D., {et~al.} 2013, Science, 341, 53

\bibitem[{{Tominaga} {et~al.}(2018){Tominaga}, {Tanaka}, {Morokuma}, {Utsumi},
  {Yamaguchi}, {Yasuda}, {Tanaka}, {Yoshida}, {Fujiyoshi}, {Furusawa},
  {Kawabata}, {Lee}, {Motohara}, {Ohsawa}, {Ohta}, {Terai}, {Abe}, {Aoki},
  {Asakura}, {Barway}, {Bond}, {Fujisawa}, {Honda}, {Ioka}, {Itoh}, {Kawai},
  {Kim}, {Koshimoto}, {Matsubayashi}, {Miyazaki}, {Saito}, {Sekiguchi}, {Sumi},
  \& {Tristram}}]{tom17gw170817}
{Tominaga}, N., {et~al.} 2018, \pasj, arXiv:1710.05865

\bibitem[{{Totani}(2013)}]{tot13}
{Totani}, T. 2013, \pasj, 65, L12

\bibitem[{{Utsumi} {et~al.}(2017){Utsumi}, {Tanaka}, {Tominaga}, {Yoshida},
  {Barway}, {Nagayama}, {Zenko}, {Aoki}, {Fujiyoshi}, {Furusawa}, {Kawabata},
  {Koshida}, {Lee}, {Morokuma}, {Motohara}, {Nakata}, {Ohsawa}, {Ohta},
  {Okita}, {Tajitsu}, {Tanaka}, {Terai}, {Yasuda}, {Abe}, {Asakura}, {Bond},
  {Miyazaki}, {Sumi}, {Tristram}, {Honda}, {Itoh}, {Itoh}, {Kawabata},
  {Morihana}, {Nagashima}, {Nakaoka}, {Ohshima}, {Takahashi}, {Takayama},
  {Aoki}, {Baar}, {Doi}, {Finet}, {Kanda}, {Kawai}, {Kim}, {Kuroda}, {Liu},
  {Matsubayashi}, {Murata}, {Nagai}, {Saito}, {Saito}, {Sako}, {Sekiguchi},
  {Tamura}, {Tanaka}, {Uemura}, \& {Yamaguchi}}]{utsumi17}
{Utsumi}, Y., {et~al.} 2017, \pasj, 69, 101

\bibitem[{{Utsumi} {et~al.}(2018){Utsumi}, {Tominaga}, {Tanaka}, {Morokuma},
  {Yoshida}, {Asakura}, {Finet}, {Furusawa}, {Kawabata}, {Liu}, {Matsubayashi},
  {Moritani}, {Motohara}, {Nakata}, {Ohta}, {Terai}, {Uemura}, \&
  {Yasuda}}]{uts17gw151226}
---. 2018, \pasj, 70, 1

\bibitem[{{Vieyro} {et~al.}(2017){Vieyro}, {Romero}, {Bosch-Ramon}, {Marcote},
  \& {del Valle}}]{vie17}
{Vieyro}, F.~L., {Romero}, G.~E., {Bosch-Ramon}, V., {Marcote}, B., \& {del
  Valle}, M.~V. 2017, \aap, 602, A64

\bibitem[{{Williams} \& {Berger}(2016)}]{wil16}
{Williams}, P.~K.~G., \& {Berger}, E. 2016, \apjl, 821, L22

\bibitem[{{Wilson} {et~al.}(2011){Wilson}, {Ferris}, {Axtens}, {Brown},
  {Davis}, {Hampson}, {Leach}, {Roberts}, {Saunders}, {Koribalski}, {Caswell},
  {Lenc}, {Stevens}, {Voronkov}, {Wieringa}, {Brooks}, {Edwards}, {Ekers},
  {Emonts}, {Hindson}, {Johnston}, {Maddison}, {Mahony}, {Malu}, {Massardi},
  {Mao}, {McConnell}, {Norris}, {Schnitzeler}, {Subrahmanyan}, {Urquhart},
  {Thompson}, \& {Wark}}]{ATCA}
{Wilson}, W.~E., {et~al.} 2011, \mnras, 416, 832

\bibitem[{{Yoshida} {et~al.}(2017){Yoshida}, {Suwa}, {Umeda}, {Shibata}, \&
  {Takahashi}}]{yos17}
{Yoshida}, T., {Suwa}, Y., {Umeda}, H., {Shibata}, M., \& {Takahashi}, K. 2017,
  \mnras, 471, 4275

\bibitem[{{Zhang}(2014)}]{zha14frb}
{Zhang}, B. 2014, \apjl, 780, L21

\end{thebibliography}

\begin{table}
  \tbl{Observations with Subaru/Hyper Suprime-Cam.}{%
  \begin{tabular}{cccccc}
      \hline
      UT & Epoch & Filter & Exposure & Seeing$^{\rm a}$ & Limiting \\
       &  &  & time & & magnitude$^{\rm b}$ \\
      &&& (s) & (arcsec) & (mag)\\ 
      \hline
   2016-01-07 & Day~8 & $g$ & 3150 & 1.44 & 26.4 \\
   2016-01-07 & Day~8 & $r$ & 4200 & 1.26 & 26.3 \\
   2016-01-07 & Day~8 & $i$ & 4200 & 0.91 & 26.5 \\
   2016-01-10 & Day~11 & $g$ & 3600 & 0.83 & 27.3 \\
   2016-01-10 & Day~11 & $r$ & 4080 & 0.68 & 27.3 \\
   2016-01-10 & Day~11 & $i$ & 3600 & 0.70 & 26.6 \\
   2016-01-13 & Day~14 & $g$ & 3600 & 1.31 & 26.5 \\
   2016-01-13 & Day~14 & $r$ & 3600 & 1.43 & 26.0 \\
   2016-01-13 & Day~14 & $i$ & 3600 & 0.81 & 26.4 \\
      \hline
   \multicolumn{6}{l}{$^{\rm a}$Full width at half maximum in the stacked images.}\\
   \multicolumn{6}{l}{$^{\rm b}$$5\sigma$ limiting magnitude with aperture diameter
   of twice of FWHM of seeing.}\\
    \end{tabular}}\label{tab:obs}
\end{table}

\begin{table}
  \tbl{Information regarding difference images.}{%
  \begin{tabular}{cccccc}
      \hline
       Epoch & Filter & PSF & Limiting & $50\%$ & $50\%$ \\
        &  & size$^{\rm a}$ & magnitude$^{\rm b}$ & completeness & completeness \\
        & & & & magnitude$^{\rm c}$ & magnitude$^{\rm d}$\\
      && (arcsec) & (mag) & (mag) & (mag)\\ 
      \hline
    Day~8 $-$ 14& $g$ & 1.44 & 26.3 & 26.6 & 26.3 \\
    Day~8 $-$ 14& $r$ & 1.44 & 26.0 & 26.2 & 26.0 \\
    Day~8 $-$ 14& $i$ & 0.91 & 26.2 & 26.3 & 26.0 \\
    Day~11 $-$ 14& $g$ & 1.31 & 26.7 & 26.9 & 26.6 \\
    Day~11 $-$ 14& $r$ & 1.43 & 26.2 & 26.4 & 26.1 \\
    Day~11 $-$ 14& $i$ & 0.81 & 26.3 & 26.3 & 26.1 \\
      \hline
   \multicolumn{6}{l}{$^{\rm a}$Full width at half maximum in the difference images.}\\
   \multicolumn{6}{l}{$^{\rm b}$ $5\sigma$ limiting magnitude with aperture diameter
   of twice of FWHM of PSF.}\\
   \multicolumn{6}{l}{$^{\rm c}$ PSF magnitude with the detection completeness of
   $50\%$}\\
   \multicolumn{6}{l}{before the source screening.}\\
   \multicolumn{6}{l}{$^{\rm d}$ PSF magnitude with the detection completeness of
   $50\%$}\\
   \multicolumn{6}{l}{after the source screening.}\\
    \end{tabular}}\label{tab:subtracted}
\end{table}

\normalsize
\begin{longtable}{ccccc}
 \caption{Candidate information.}\label{tab:cand}
 \hline \hline
 \multicolumn{5}{c}{Name} \\ \hline
 Separation$^{\rm a}$ & Separation$^{\rm b}$ & Radio S/N$^{\rm c}$ & RA &Dec \\ 
 (arcmin) & (arcsec) & & & \\ \hline
 Epoch & \multicolumn{2}{c}{$g$} & $r$ & $i$ \\
 & \multicolumn{2}{c}{($\upmu$Jy)} & ($\upmu$Jy) & ($\upmu$Jy) \\
 \hline
 \endfirsthead
 \hline \hline
 Name & Separation$^{\rm a}$ & Separation$^{\rm b}$ & RA &Dec \\ 
 & (arcmin) & (arcsec) & & \\ \hline
 Epoch & \multicolumn{2}{c}{$g$} & $r$ & $i$ \\
 & \multicolumn{2}{c}{($\upmu$Jy)} & ($\upmu$Jy) & ($\upmu$Jy) \\
 \hline
 \endhead
 \endfoot
 \multicolumn{5}{l}{$^{\rm a}$Separation from the center of the
 localization area of \FRB.}\\
 \multicolumn{5}{l}{$^{\rm b}$Separation from the objects in the stacked images or
 the objects in the PS1 catalog.}\\
 \multicolumn{5}{l}{$^{\rm c}$Intrinsic radio S/N of \FRB\ at the candidate location.}\\
 \multicolumn{5}{l}{$^{\rm d}$Upper limits for the fluxes of the candidates.}\\
 \endlastfoot
\hline \hline
 \multicolumn{5}{c}{Cand-1} \\ \hline
 $13.92$ & $0.12$ & $250$ & 09:40:10.67 & $-$03:36:56.4\\ \hline
Day~8 & \multicolumn{2}{c}{$+9.515\pm0.015^{\rm d}$} & $+16.574\pm0.019^{\rm d}$ & $+15.717\pm0.017^{\rm d}$\\
Day~11 & \multicolumn{2}{c}{$+5.607\pm0.009^{\rm d}$} & $+8.410\pm0.010^{\rm d}$ & $+11.845\pm0.016^{\rm d}$\\
Day~14 & \multicolumn{2}{c}{$+8.760\pm0.012^{\rm d}$} & $+17.617\pm0.019^{\rm d}$ & $+13.405\pm0.017^{\rm d}$\\
Day~8 $-$ Day~14 & \multicolumn{2}{c}{$-0.167\pm0.016$} & $+0.070\pm0.019$ & $<+0.055$\\
Day~11 $-$ Day~14 & \multicolumn{2}{c}{$-0.108\pm0.012$} & $<+0.054$ & $-0.107\pm0.018$\\
 \hline \hline 
 \multicolumn{5}{c}{Cand-2} \\ \hline
 $6.24$ & $1.05$ & $29$ & 09:40:29.18 & $-$03:30:32.2\\ \hline
Day~8 & \multicolumn{2}{c}{$+0.454\pm0.012^{\rm d}$} & $+0.559\pm0.015^{\rm d}$ & $+0.605\pm0.014^{\rm d}$\\
Day~11 & \multicolumn{2}{c}{$+0.450\pm0.007^{\rm d}$} & $+0.460\pm0.007^{\rm d}$ & $+0.556\pm0.013^{\rm d}$\\
Day~14 & \multicolumn{2}{c}{$+0.596\pm0.010^{\rm d}$} & $+0.717\pm0.015^{\rm d}$ & $+0.697\pm0.016^{\rm d}$\\
Day~8 $-$ Day~14 & \multicolumn{2}{c}{$-0.160\pm0.013$} & $-0.114\pm0.016$ & $-0.167\pm0.015$\\
Day~11 $-$ Day~14 & \multicolumn{2}{c}{$-0.056\pm0.010$} & $<+0.045$ & $-0.090\pm0.017$\\
 \hline \hline 
 \multicolumn{5}{c}{Cand-3} \\ \hline
 $10.15$ & $0.43$ & $71$ & 09:40:11.07 & $-$03:29:59.4\\ \hline
Day~8 & \multicolumn{2}{c}{$+0.486\pm0.014^{\rm d}$} & $+0.507\pm0.016^{\rm d}$ & $+0.483\pm0.015^{\rm d}$\\
Day~11 & \multicolumn{2}{c}{$+0.257\pm0.007^{\rm d}$} & $+0.298\pm0.008^{\rm d}$ & $+0.393\pm0.014^{\rm d}$\\
Day~14 & \multicolumn{2}{c}{$+0.242\pm0.010^{\rm d}$} & $+0.383\pm0.016^{\rm d}$ & $+0.344\pm0.015^{\rm d}$\\
Day~8 $-$ Day~14 & \multicolumn{2}{c}{$+0.212\pm0.014$} & $+0.173\pm0.016$ & $+0.106\pm0.016$\\
Day~11 $-$ Day~14 & \multicolumn{2}{c}{$+0.107\pm0.010$} & $+0.111\pm0.015$ & $+0.080\pm0.016$\\
 \hline \hline 
 \multicolumn{5}{c}{Cand-4} \\ \hline
 $10.08$ & $0.05$ & $70$ & 09:40:10.39 & $-$03:29:02.1\\ \hline
Day~8 & \multicolumn{2}{c}{$+4.365\pm0.015^{\rm d}$} & $+11.005\pm0.017^{\rm d}$ & $+13.500\pm0.016^{\rm d}$\\
Day~11 & \multicolumn{2}{c}{$+2.552\pm0.008^{\rm d}$} & $+6.310\pm0.010^{\rm d}$ & $+10.800\pm0.015^{\rm d}$\\
Day~14 & \multicolumn{2}{c}{$+3.897\pm0.012^{\rm d}$} & $+11.718\pm0.017^{\rm d}$ & $+11.945\pm0.018^{\rm d}$\\
Day~8 $-$ Day~14 & \multicolumn{2}{c}{$<+0.046$} & $-0.077\pm0.018$ & $-0.173\pm0.017$\\
Day~11 $-$ Day~14 & \multicolumn{2}{c}{$<+0.035$} & $-0.059\pm0.017$ & $-0.249\pm0.019$\\
 \hline \hline 
 \multicolumn{5}{c}{Cand-5} \\ \hline
 $13.04$ & $0.13$ &$180$& 09:39:57.85 & $-$03:26:25.3\\ \hline
Day~8 & \multicolumn{2}{c}{$+8.552\pm0.015^{\rm d}$} & $+19.262\pm0.018^{\rm d}$ & $+24.074\pm0.018^{\rm d}$\\
Day~11 & \multicolumn{2}{c}{$+6.213\pm0.009^{\rm d}$} & $+12.303\pm0.011^{\rm d}$ & $+20.053\pm0.017^{\rm d}$\\
Day~14 & \multicolumn{2}{c}{$+8.361\pm0.013^{\rm d}$} & $+20.372\pm0.020^{\rm d}$ & $+22.062\pm0.019^{\rm d}$\\
Day~8 $-$ Day~14 & \multicolumn{2}{c}{$<+0.046$} & $-0.491\pm0.021$ & $-0.252\pm0.019$\\
Day~11 $-$ Day~14 & \multicolumn{2}{c}{$-0.066\pm0.013$} & $-0.226\pm0.020$ & $-0.401\pm0.021$\\
 \hline \hline 
 \multicolumn{5}{c}{Cand-6} \\ \hline
 $11.43$ & $0.15$ &$100$& 09:40:11.61 & $-$03:20:51.9\\ \hline
Day~8 & \multicolumn{2}{c}{$+1.308\pm0.014^{\rm d}$} & $+1.108\pm0.016^{\rm d}$ & $+1.035\pm0.014^{\rm d}$\\
Day~11 & \multicolumn{2}{c}{$+0.926\pm0.008^{\rm d}$} & $+0.790\pm0.008^{\rm d}$ & $+0.910\pm0.013^{\rm d}$\\
Day~14 & \multicolumn{2}{c}{$+1.387\pm0.011^{\rm d}$} & $+1.452\pm0.017^{\rm d}$ & $+1.152\pm0.016^{\rm d}$\\
Day~8 $-$ Day~14 & \multicolumn{2}{c}{$-0.243\pm0.014$} & $-0.197\pm0.017$ & $-0.240\pm0.015$\\
Day~11 $-$ Day~14 & \multicolumn{2}{c}{$-0.086\pm0.011$} & $<+0.050$ & $-0.136\pm0.017$\\
 \hline \hline 
 \multicolumn{5}{c}{Cand-7} \\ \hline
 $8.44$ & $0.01$ &$46$& 09:40:19.25 & $-$03:23:35.5\\ \hline
Day~8 & \multicolumn{2}{c}{$+9.660\pm0.015^{\rm d}$} & $+20.887\pm0.018^{\rm d}$ & $+27.563\pm0.018^{\rm d}$\\
Day~11 & \multicolumn{2}{c}{$+6.936\pm0.009^{\rm d}$} & $+13.976\pm0.011^{\rm d}$ & $+23.124\pm0.017^{\rm d}$\\
Day~14 & \multicolumn{2}{c}{$+8.744\pm0.013^{\rm d}$} & $+21.718\pm0.020^{\rm d}$ & $+25.308\pm0.020^{\rm d}$\\
Day~8 $-$ Day~14 & \multicolumn{2}{c}{$+0.439\pm0.016$} & $+0.414\pm0.020$ & $+0.114\pm0.019$\\
Day~11 $-$ Day~14 & \multicolumn{2}{c}{$+0.066\pm0.013$} & $+0.102\pm0.019$ & $-0.380\pm0.021$\\
 \hline \hline 
 \multicolumn{5}{c}{Cand-8} \\ \hline
 $8.68$ & $0.03$ &$48$& 09:40:18.22 & $-$03:23:35.1\\ \hline
Day~8 & \multicolumn{2}{c}{$+9.394\pm0.016^{\rm d}$} & $+26.611\pm0.020^{\rm d}$ & $+37.181\pm0.020^{\rm d}$\\
Day~11 & \multicolumn{2}{c}{$+6.789\pm0.009^{\rm d}$} & $+18.755\pm0.013^{\rm d}$ & $+31.848\pm0.020^{\rm d}$\\
Day~14 & \multicolumn{2}{c}{$+8.917\pm0.013^{\rm d}$} & $+28.091\pm0.021^{\rm d}$ & $+34.622\pm0.023^{\rm d}$\\
Day~8 $-$ Day~14 & \multicolumn{2}{c}{$+0.077\pm0.016$} & $<+0.064$ & $-0.625\pm0.021$\\
Day~11 $-$ Day~14 & \multicolumn{2}{c}{$+0.094\pm0.013$} & $+0.281\pm0.020$ & $-0.731\pm0.024$\\
 \hline \hline 
 \multicolumn{5}{c}{Cand-9} \\ \hline
 $12.54$ & $0.10$ &$150$& 09:40:02.18 & $-$03:23:16.7\\ \hline
Day~8 & \multicolumn{2}{c}{$+5.229\pm0.015^{\rm d}$} & $+12.062\pm0.018^{\rm d}$ & $+15.041\pm0.016^{\rm d}$\\
Day~11 & \multicolumn{2}{c}{$+3.852\pm0.008^{\rm d}$} & $+8.492\pm0.010^{\rm d}$ & $+12.662\pm0.015^{\rm d}$\\
Day~14 & \multicolumn{2}{c}{$+4.915\pm0.012^{\rm d}$} & $+12.842\pm0.018^{\rm d}$ & $+13.898\pm0.018^{\rm d}$\\
Day~8 $-$ Day~14 & \multicolumn{2}{c}{$<+0.045$} & $+0.223\pm0.018$ & $+0.191\pm0.017$\\
Day~11 $-$ Day~14 & \multicolumn{2}{c}{$<+0.037$} & $-0.067\pm0.018$ & $+0.132\pm0.019$\\
 \hline \hline 
 \multicolumn{5}{c}{Cand-10} \\ \hline
 $10.59$ & $0.03$ &$81$& 09:41:23.97 & $-$03:20:44.9\\ \hline
Day~8 & \multicolumn{2}{c}{$+7.762\pm0.015^{\rm d}$} & $+12.770\pm0.017^{\rm d}$ & $+14.591\pm0.016^{\rm d}$\\
Day~11 & \multicolumn{2}{c}{$+5.266\pm0.008^{\rm d}$} & $+8.069\pm0.010^{\rm d}$ & $+11.342\pm0.014^{\rm d}$\\
Day~14 & \multicolumn{2}{c}{$+7.222\pm0.012^{\rm d}$} & $+14.180\pm0.018^{\rm d}$ & $+12.727\pm0.017^{\rm d}$\\
Day~8 $-$ Day~14 & \multicolumn{2}{c}{$+0.178\pm0.015$} & $-0.380\pm0.018$ & $-0.350\pm0.017$\\
Day~11 $-$ Day~14 & \multicolumn{2}{c}{$<+0.036$} & $-0.275\pm0.018$ & $-0.198\pm0.018$\\
 \hline \hline 
 \multicolumn{5}{c}{Cand-11} \\ \hline
 $10.55$ & $0.06$ &$80$& 09:41:26.72 & $-$03:21:52.2\\ \hline
Day~8 & \multicolumn{2}{c}{$+7.634\pm0.014^{\rm d}$} & $+11.339\pm0.016^{\rm d}$ & $+11.362\pm0.015^{\rm d}$\\
Day~11 & \multicolumn{2}{c}{$+4.762\pm0.008^{\rm d}$} & $+6.245\pm0.009^{\rm d}$ & $+8.653\pm0.014^{\rm d}$\\
Day~14 & \multicolumn{2}{c}{$+7.136\pm0.012^{\rm d}$} & $+12.607\pm0.017^{\rm d}$ & $+10.132\pm0.017^{\rm d}$\\
Day~8 $-$ Day~14 & \multicolumn{2}{c}{$+0.071\pm0.015$} & $-0.084\pm0.018$ & $-0.227\pm0.016$\\
Day~11 $-$ Day~14 & \multicolumn{2}{c}{$<+0.037$} & $<+0.051$ & $-0.188\pm0.018$\\
 \hline \hline 
 \multicolumn{5}{c}{Cand-12} \\ \hline
 $10.68$ & $0.24$ &$83$& 09:41:12.99 & $-$03:18:04.8\\ \hline
Day~8 & \multicolumn{2}{c}{$+17.315\pm0.017^{\rm d}$} & $+32.989\pm0.020^{\rm d}$ & $+37.663\pm0.020^{\rm d}$\\
Day~11 & \multicolumn{2}{c}{$+10.673\pm0.010^{\rm d}$} & $+19.520\pm0.012^{\rm d}$ & $+28.782\pm0.018^{\rm d}$\\
Day~14 & \multicolumn{2}{c}{$+14.755\pm0.013^{\rm d}$} & $+33.926\pm0.020^{\rm d}$ & $+31.625\pm0.021^{\rm d}$\\
Day~8 $-$ Day~14 & \multicolumn{2}{c}{$+1.338\pm0.017$} & $+0.744\pm0.021$ & $+2.011\pm0.021$\\
Day~11 $-$ Day~14 & \multicolumn{2}{c}{$+0.549\pm0.013$} & $+0.270\pm0.020$ & $+0.969\pm0.022$\\
 \hline \hline 
 \multicolumn{5}{c}{Cand-13} \\ \hline
 $8.84$ & $0.19$ &$50$& 09:40:30.93 & $-$03:19:38.0\\ \hline
Day~8 & \multicolumn{2}{c}{$+8.246\pm0.014^{\rm d}$} & $+25.759\pm0.018^{\rm d}$ & $+37.745\pm0.020^{\rm d}$\\
Day~11 & \multicolumn{2}{c}{$+4.469\pm0.008^{\rm d}$} & $+14.175\pm0.011^{\rm d}$ & $+30.174\pm0.018^{\rm d}$\\
Day~14 & \multicolumn{2}{c}{$+7.466\pm0.012^{\rm d}$} & $+28.891\pm0.020^{\rm d}$ & $+34.284\pm0.021^{\rm d}$\\
Day~8 $-$ Day~14 & \multicolumn{2}{c}{$<+0.044$} & $-0.117\pm0.020$ & $-0.250\pm0.021$\\
Day~11 $-$ Day~14 & \multicolumn{2}{c}{$<+0.035$} & $-0.170\pm0.019$ & $-0.542\pm0.022$\\
 \hline
\end{longtable}

\begin{longtable}{ccccccc}
 \caption{Templates fitting the difference light curves of candidates.}\label{tab:model}
 \hline \hline
   \multicolumn{7}{c}{Name (photometric redshift)} \\ \hline
   SN Type & Number$^{\rm a}$ & 
   Redshift & $\texp$& 
   $\Ebvh$ & $\MB$ &  
   $Q$ value\\
 \hline
 \endfirsthead
 \hline \hline
   \multicolumn{7}{c}{Name (photometric redshift)} \\ \hline
   SN Type & Number & 
   Redshift & $\texp$& 
   $\Ebvh$ & $\MB$ &  
   $Q$ value\\
 \hline
 \endhead
 \endfoot
   \multicolumn{7}{l}{$^{\rm a}$The number of templates fitting the
 multicolor light curves of candidates.}\\
   \multicolumn{7}{l}{$^{\rm b}$The best-fit template.}\\
 \endlastfoot
 \hline \hline
\multicolumn{7}{c}{Cand-2 (photo-$z=0.18_{-0.15}^{+0.46}$)} \\ \hline
PS1-10ah$^{\rm b}$ & 30 & $0.60_{-0.23}^{+0.03}$  & $+7.2_{-0.5}^{+1.1}$  & $0.00_{-0.00}^{+0.24}$  & $-16.78_{-0.13}^{+0.18}$  & $2.1\times10^{-4}-2.1\times10^{-2}$\\
PS1-10bjp & 34 & $0.55_{-0.18}^{+0.08}$  & $+7.3_{-0.6}^{+1.0}$  & $0.20_{-0.20}^{+0.23}$  & $-17.20_{-0.03}^{+0.03}$  & $2.5\times10^{-4}-1.4\times10^{-2}$\\
Nugent IIn & 116 & $0.55_{-0.23}^{+0.08}$  & $+3.9_{-4.2}^{+2.1}$  & $0.14_{-0.14}^{+0.31}$  & $-16.96_{-0.38}^{+0.03}$  & $2.2\times10^{-4}-1.0\times10^{-2}$\\
Nugent IIP & 1 & $0.20_{-0.03}^{+0.03}$  & $+5.0_{-0.5}^{+0.5}$  & $0.05_{-0.05}^{+0.01}$  & $-14.41_{-0.03}^{+0.03}$  & $2.1\times10^{-4}$\\
\hline \hline \multicolumn{7}{c}{Cand-3} \\ \hline
PS1-11qr$^{\rm b}$ & 9 & $0.60_{-0.13}^{+0.03}$  & $-12.8_{-0.5}^{+2.8}$  & $0.00_{-0.00}^{+0.11}$  & $-18.13_{-0.08}^{+0.13}$  & $2.2\times10^{-8}-2.2\times10^{-6}$\\
PS1-12brf & 1 & $0.65_{-0.03}^{+0.03}$  & $-2.5_{-0.5}^{+0.5}$  & $0.00_{-0.00}^{+0.01}$  & $-17.94_{-0.03}^{+0.03}$  & $2.3\times10^{-8}$\\
\hline \hline \multicolumn{7}{c}{Cand-6 (photo-$z=1.44_{-0.27}^{+0.46}$)} \\ \hline
Nugent IIn$^{\rm b}$ & 29 & $1.25_{-0.08}^{+0.48}$  & $+0.8_{-2.4}^{+1.1}$  & $0.00_{-0.00}^{+0.01}$  & $-18.71_{-0.78}^{+0.23}$  & $6.7\times10^{-6}-2.5\times10^{-4}$\\
\hline \hline \multicolumn{7}{c}{Cand-11 (photo-$z=0.19_{-0.15}^{+0.07}$)} \\ \hline
Nugent IIn$^{\rm b}$ & 8 & $0.10_{-0.03}^{+0.03}$  & $-6.3_{-0.5}^{+0.5}$  & $0.57_{-0.57}^{+0.07}$  & $-16.96_{-0.03}^{+0.03}$  & $1.5\times10^{-4}-3.0\times10^{-3}$\\
 \hline
\end{longtable}

\begin{longtable}{ccccccc}
 \caption{Templates fitting the difference light curve of \SNb\
 stretching the RT templates.}\label{tab:modelSNb}
 \hline \hline
   \multicolumn{7}{c}{Stretch} \\ \hline
   SN Type & Number$^{\rm a}$ & 
   Redshift & $\texp$& 
   $\Ebvh$ & $\MB$ &  
   $Q$ value\\
 \hline
 \endfirsthead
 \hline \hline
   \multicolumn{7}{c}{Stretch} \\ \hline
   SN Type & Number & 
   Redshift & $\texp$& 
   $\Ebvh$ & $\MB$ &  
   $Q$ value\\
 \hline
 \endhead
 \endfoot
   \multicolumn{7}{l}{$^{\rm a}$The number of templates fitting the
 multicolor light curves of candidates.}\\
   \multicolumn{7}{l}{$^{\rm b}$The best-fit template.}\\
 \endlastfoot
\hline \hline \multicolumn{7}{c}{$s=0.8$} \\ \hline
PS1-10ah$^{\rm b}$ & 45 & $0.20_{-0.03}^{+0.03}$ & $-1.4_{-2.5}^{+7.5}$  &  $0.20_{-0.16}^{+0.18}$  & $-16.63_{-0.03}^{+0.03}$  & $7.4\times10^{-6}-7.4\times10^{-4}$\\
PS1-12brf & 4 & $0.65_{-0.03}^{+0.23}$ & $+0.8_{-1.5}^{+0.5}$  &  $0.00_{-0.01}^{+0.01}$  & $-17.94_{-0.63}^{+0.08}$  & $9.2\times10^{-6}-5.5\times10^{-4}$\\
PS1-10bjp & 3 & $0.40_{-0.03}^{+0.03}$ & $+3.6_{-0.5}^{+0.5}$  &  $0.16_{-0.02}^{+0.02}$  & $-17.20_{-0.03}^{+0.03}$  & $2.7\times10^{-5}-3.2\times10^{-4}$\\
\hline \hline \multicolumn{7}{c}{$s=0.6$} \\ \hline
PS1-10bjp$^{\rm b}$ & 11 & $0.40_{-0.03}^{+0.13}$ & $+5.2_{-0.5}^{+0.5}$  &  $0.16_{-0.12}^{+0.03}$  & $-17.20_{-0.03}^{+0.03}$  & $6.1\times10^{-3}-4.2\times10^{-1}$\\
PS1-10ah & 48 & $0.35_{-0.03}^{+0.18}$ & $+5.2_{-7.5}^{+1.5}$  &  $0.06_{-0.07}^{+0.06}$  & $-16.63_{-1.48}^{+0.03}$  & $4.3\times10^{-3}-3.0\times10^{-2}$\\
 \hline
\end{longtable}

\scriptsize
\begin{longtable}{p{1.2cm}p{1.3cm}p{1.1cm}p{1.5cm}p{1.cm}p{1.5cm}p{0.5cm}p{0.5cm}p{0.5cm}p{1.2cm}p{1.2cm}p{1.2cm}}
 \caption{Properties of candidates}\label{tab:scores}
 \hline \hline
 Name & Likely origin & Location within galaxies & Probable transient templates &
 Photo-$z$ consistent with DM & Consistent with non-detection in the
 other beams & ATCA source & VLA source & GMRT source & AGN signature
 in WISE color & Low-mass host galaxy &
 Star-forming host galaxy\\ 
 \hline
 \endfirsthead
 \endhead
 \endfoot
 \multicolumn{12}{l}{$^{\rm a}$The fluxes of host galaxy cannot be
 derived due to the non-negligible contribution of the candidate to the
 total flux.}\\
 \multicolumn{12}{l}{$^{\rm b}$No WISE source is found within the
 spatial resolution element of WISE.}\\
 \multicolumn{12}{l}{$^{\rm c}$This excludes the candidate as an optical
 counterpart to \FRB.}\\
 \endlastfoot
 \hline
 Cand-1  & AGN & Center & None & Yes &Yes  & Yes & No & No & No & No & Yes \\
 Cand-2  & CCSN/RT & Off-center & RT and SNe~IIP and IIn & Yes  &Yes  & No & No  & No &
 Yes, but there are multiple galaxies & Yes & Yes  \\
 Cand-3  & RT & Off-center & None, but reproduced with RTs with a faint peak and rapid decline& ---$^{\rm a}$  &Yes  &No & No & No  & ---$^{\rm b}$  & ---$^{\rm a}$ & ---$^{\rm a}$\\
 Cand-4  & AGN & Center & None & Yes  & Yes  & No & No& No & No & No & Yes    \\
 Cand-5  & AGN & Center & None & Yes & {\bf No}$^{\rm c}$  & No & No & No & No  & No & No  \\
 Cand-6  & CCSN/AGN & Center & SN~IIn & {\bf No}$^{\rm c}$  &Yes   & No & No & No & ---$^{\rm b}$ & No & Yes    \\
 Cand-7  & AGN & Center & None & Yes  & Yes & No & No & No & No  & No & Yes   \\
 Cand-8  & AGN & Center & None & Yes  & Yes & No &No  & No & No  & No & Yes   \\
 Cand-9  & AGN & Center & None & Yes  & Yes &No  & No & No & ---$^{\rm b}$ & No & Yes    \\
 Cand-10 & AGN & Center & None & Yes  & Yes &Yes  & No & No & No & No & Yes    \\
 Cand-11 & CCSN/AGN & Center & SN~IIn &Yes   & Yes & No & No & No & ---$^{\rm b}$ & No & Yes    \\
 Cand-12 & AGN & Center & None & Yes  & Yes & No & No& No & No   & No & Yes   \\
 Cand-13 & AGN & Center & None & Yes  & Yes & No & No& No & No  & No & Yes   \\
\hline
\end{longtable}
\normalsize

\scriptsize
\begin{longtable}{cccccccc}
 \caption{Templates fitting the light curves of transients outside
 the localization area of \FRB.}\label{tab:model3}
 \hline \hline
 \multicolumn{8}{c}{Name (separation$^{\rm a}$, photometric redshift)} \\ \hline
 SN Type & Number$^{\rm b}$ & Redshift & $\texp$& 
   \multicolumn{2}{c}{$\Ebvh$$^{\rm c}$} &$\MB$ &  
   $Q$ value\\
   &&&&Stretch$^{\rm d}$ & Color$^{\rm d}$ &  & \\
 \hline
 \endfirsthead
 \hline \hline
 \multicolumn{8}{c}{Name (separation$^{\rm a}$, photometric redshift)} \\ \hline
 SN Type & Number$^{\rm b}$ & Redshift & $\texp$& 
   \multicolumn{2}{c}{$\Ebvh$$^{\rm c}$} &$\MB$ &  
   $Q$ value\\
   &&&&Stretch$^{\rm d}$ & Color$^{\rm d}$ &  & \\
 \hline
 \endhead
 \endfoot
 \multicolumn{8}{l}{$^{\rm a}$Separation from the center of the
 localization area of \FRB.}\\
 \multicolumn{8}{l}{$^{\rm b}$The number of templates fitting the
 multicolor light curves of candidates.}\\
 \multicolumn{8}{l}{$^{\rm c}$Color excess of the host galaxy for the CCSN and RT templates.}\\
 \multicolumn{8}{l}{$^{\rm d}$Stretch and color parameters for the SN~Ia templates.}\\
 \multicolumn{8}{l}{$^{\rm e}$The best-fit template.}\\
 \endlastfoot
 \hline
\multicolumn{8}{c}{Cand-14 ($44.91$~arcmin, photo-$z=0.21^{+0.07}_{-0.16}$)} \\ \hline
Nugent Ibc normal$^{\rm e}$ & 16 & $0.20_{-0.08}^{+0.08}$  & $-12.2_{-2.3}^{+0.5}$  & \multicolumn{2}{c}{$0.05_{-0.05}^{+0.15}$ } & $-16.05_{-0.33}^{+0.03}$  & $1.0\times10^{-3}-2.9\times10^{-3}$\\
\hline \hline \multicolumn{8}{c}{Cand-15 ($41.17$~arcmin, photo-$z=0.55^{+0.06}_{-0.05}$)} \\ \hline
Hsiao Ia$^{\rm e}$ & 280 & $0.60_{-0.13}^{+0.03}$  & $-14.5_{-2.5}^{+19.0}$  & $1.20_{-0.56}^{+0.01}$  & $0.65_{-0.13}^{+0.13}$  & $-18.08_{-0.55}^{+0.94}$  & $2.4\times10^{-3}-6.3\times10^{-2}$\\
Nugent Ibc bright & 102 & $0.50_{-0.03}^{+0.13}$  & $+4.0_{-4.3}^{+2.5}$  & \multicolumn{2}{c}{$0.26_{-0.26}^{+0.07}$ } & $-18.45_{-0.03}^{+0.03}$  & $2.5\times10^{-3}-5.2\times10^{-2}$\\
Nugent Ibc normal & 27 & $0.60_{-0.13}^{+0.03}$  & $+3.4_{-1.5}^{+3.1}$  & \multicolumn{2}{c}{$0.00_{-0.00}^{+0.01}$ } & $-18.05_{-0.03}^{+0.63}$  & $2.6\times10^{-3}-1.4\times10^{-2}$\\
Nugent IIL normal & 19 & $0.55_{-0.08}^{+0.08}$  & $-2.5_{-1.0}^{+1.0}$  & \multicolumn{2}{c}{$0.20_{-0.20}^{+0.11}$ } & $-17.91_{-0.03}^{+0.80}$  & $2.4\times10^{-3}-9.0\times10^{-3}$\\
\hline \hline \multicolumn{8}{c}{Cand-16 ($28.41$~arcmin, photo-$z=0.47^{+0.14}_{-0.14}$)} \\ \hline
Hsiao Ia$^{\rm e}$ & 18 & $0.45_{-0.03}^{+0.03}$  & $-14.7_{-2.9}^{+8.1}$  & $1.15_{-0.36}^{+0.06}$  & $0.75_{-0.18}^{+0.03}$  & $-17.79_{-0.73}^{+0.35}$  & $9.1\times10^{-3}-1.3\times10^{-1}$\\
Nugent IIP & 12 & $0.55_{-0.08}^{+0.08}$  & $-2.5_{-1.0}^{+4.0}$  & \multicolumn{2}{c}{$0.00_{-0.00}^{+0.01}$ } & $-18.01_{-0.43}^{+0.33}$  & $9.2\times10^{-3}-1.1\times10^{-1}$\\
Nugent IIL normal & 17 & $0.60_{-0.13}^{+0.03}$  & $-3.0_{-0.5}^{+1.5}$  & \multicolumn{2}{c}{$0.04_{-0.04}^{+0.17}$ } & $-17.91_{-0.03}^{+0.50}$  & $9.0\times10^{-3}-8.7\times10^{-2}$\\
Nugent Ibc normal & 1 & $0.40_{-0.03}^{+0.03}$  & $-7.4_{-0.5}^{+0.5}$  & \multicolumn{2}{c}{$0.00_{-0.00}^{+0.01}$ } & $-17.90_{-0.03}^{+0.03}$  & $9.0\times10^{-3}$\\
\hline \hline \multicolumn{8}{c}{Cand-17 ($46.15$~arcmin, photo-$z=0.33^{+0.22}_{-0.15}$)} \\ \hline
Nugent Ibc bright$^{\rm e}$ & 75 & $0.50_{-0.08}^{+0.08}$  & $-30.0_{-2.5}^{+7.3}$  & \multicolumn{2}{c}{$0.01_{-0.01}^{+0.14}$ } & $-18.45_{-0.38}^{+0.03}$  & $1.0\times10^{-2}-9.7\times10^{-1}$\\
Hsiao Ia & 14162 & $0.55_{-0.38}^{+0.03}$  & $-49.3_{-27.7}^{+35.9}$  & $1.20_{-0.56}^{+0.01}$  & $0.30_{-0.53}^{+0.48}$  & $-19.17_{-1.12}^{+2.03}$  & $9.7\times10^{-3}-8.7\times10^{-1}$\\
Nugent Ibc normal & 160 & $0.20_{-0.03}^{+0.28}$  & $-33.2_{-5.3}^{+10.5}$  & \multicolumn{2}{c}{$0.00_{-0.00}^{+0.01}$ } & $-16.70_{-1.38}^{+0.63}$  & $1.0\times10^{-2}-8.6\times10^{-1}$\\
Nugent IIL bright & 186 & $0.40_{-0.03}^{+0.18}$  & $-25.0_{-13.5}^{+18.0}$  & \multicolumn{2}{c}{$0.00_{-0.00}^{+0.01}$ } & $-18.86_{-0.63}^{+0.78}$  & $9.8\times10^{-3}-1.4\times10^{-1}$\\
Nugent IIP & 45 & $0.40_{-0.03}^{+0.18}$  & $-27.0_{-6.5}^{+20.0}$  & \multicolumn{2}{c}{$0.00_{-0.00}^{+0.01}$ } & $-18.91_{-0.03}^{+0.38}$  & $1.1\times10^{-2}-1.2\times10^{-1}$\\
\hline \hline \multicolumn{8}{c}{Cand-18 ($28.28$~arcmin, $-$)} \\ \hline
Hsiao Ia$^{\rm e}$ & 65 & $0.30_{-0.08}^{+0.13}$  & $-51.2_{-6.1}^{+20.1}$  & $1.10_{-0.36}^{+0.11}$  & $-0.15_{-0.08}^{+0.13}$  & $-19.78_{-0.26}^{+0.70}$  & $3.4\times10^{-3}-3.3\times10^{-1}$\\
Nugent IIL bright & 1 & $0.35_{-0.03}^{+0.03}$  & $-8.5_{-0.5}^{+0.5}$  & \multicolumn{2}{c}{$0.00_{-0.00}^{+0.01}$ } & $-18.11_{-0.03}^{+0.03}$  & $4.0\times10^{-3}$\\
\hline \hline \multicolumn{8}{c}{Cand-19 ($29.12$~arcmin, photo-$z=0.88^{+0.11}_{-0.06}$)} \\ \hline
Nugent IIn$^{\rm e}$ & 5 & $0.90_{-0.03}^{+0.08}$  & $+2.3_{-1.2}^{+0.5}$  & \multicolumn{2}{c}{$0.00_{-0.00}^{+0.01}$ } & $-18.21_{-0.28}^{+0.08}$  & $6.7\times10^{-4}-5.7\times10^{-3}$\\
\hline \hline \multicolumn{8}{c}{Cand-20 ($22.71$~arcmin, $-$)} \\ \hline
Hsiao Ia$^{\rm e}$ & 63 & $0.60_{-0.18}^{+0.03}$  & $+9.2_{-1.1}^{+1.4}$  & $0.65_{-0.01}^{+0.56}$  & $-0.20_{-0.03}^{+0.23}$  & $-19.63_{-0.71}^{+0.78}$  & $2.9\times10^{-3}-2.9\times10^{-1}$\\
\hline \hline \multicolumn{8}{c}{Cand-21 ($32.90$~arcmin, photo-$z=0.28^{+0.32}_{-0.20}$)} \\ \hline
Hsiao Ia$^{\rm e}$ & 22 & $0.35_{-0.18}^{+0.03}$  & $+10.8_{-1.2}^{+0.5}$  & $0.75_{-0.11}^{+0.46}$  & $-0.20_{-0.03}^{+0.18}$  & $-19.46_{-0.82}^{+0.42}$  & $2.0\times10^{-3}-2.0\times10^{-1}$\\
\hline \hline \multicolumn{8}{c}{Cand-22 ($37.47$~arcmin, photo-$z=0.45^{+0.13}_{-0.26}$)} \\ \hline
Hsiao Ia$^{\rm e}$ & 274 & $0.55_{-0.13}^{+0.03}$  & $+3.7_{-10.4}^{+2.9}$  & $1.20_{-0.51}^{+0.01}$  & $-0.15_{-0.08}^{+0.18}$  & $-20.05_{-0.29}^{+1.02}$  & $1.3\times10^{-5}-1.3\times10^{-3}$\\
\hline \hline \multicolumn{8}{c}{Cand-23 ($46.41$~arcmin, $-$)} \\ \hline
Nugent Ibc bright$^{\rm e}$ & 77 & $0.65_{-0.23}^{+0.08}$  & $-0.4_{-3.3}^{+3.9}$  & \multicolumn{2}{c}{$0.00_{-0.00}^{+0.26}$ } & $-18.80_{-0.38}^{+0.38}$  & $6.7\times10^{-3}-6.6\times10^{-1}$\\
Hsiao Ia & 123 & $0.65_{-0.18}^{+0.28}$  & $-3.1_{-4.4}^{+4.8}$  & $0.80_{-0.16}^{+0.41}$  & $0.30_{-0.53}^{+0.43}$  & $-18.38_{-1.96}^{+0.95}$  & $6.7\times10^{-3}-2.0\times10^{-1}$\\
Nugent IIL normal & 11 & $0.40_{-0.03}^{+0.13}$  & $-1.0_{-1.5}^{+0.5}$  & \multicolumn{2}{c}{$0.00_{-0.00}^{+0.32}$ } & $-16.89_{-1.05}^{+0.08}$  & $7.0\times10^{-3}-5.6\times10^{-2}$\\
\hline \hline \multicolumn{8}{c}{Cand-24 ($38.75$~arcmin, photo-$z=0.56^{+0.06}_{-0.33}$)} \\ \hline
Nugent Ibc normal$^{\rm e}$ & 158 & $0.35_{-0.13}^{+0.23}$  & $+4.4_{-8.3}^{+3.1}$  & \multicolumn{2}{c}{$0.00_{-0.00}^{+0.15}$ } & $-16.65_{-1.33}^{+0.63}$  & $9.4\times10^{-3}-8.6\times10^{-1}$\\
Hsiao Ia & 862 & $0.60_{-0.18}^{+0.03}$  & $-0.5_{-1.5}^{+9.1}$  & $1.20_{-0.56}^{+0.01}$  & $0.20_{-0.23}^{+0.43}$  & $-19.00_{-0.61}^{+1.29}$  & $8.6\times10^{-3}-5.6\times10^{-1}$\\
Nugent IIP & 24 & $0.25_{-0.03}^{+0.28}$  & $+1.5_{-1.5}^{+4.0}$  & \multicolumn{2}{c}{$0.00_{-0.00}^{+0.01}$ } & $-15.41_{-1.93}^{+0.13}$  & $8.8\times10^{-3}-2.8\times10^{-1}$\\
Nugent IIL normal & 97 & $0.30_{-0.08}^{+0.23}$  & $+0.0_{-1.0}^{+5.5}$  & \multicolumn{2}{c}{$0.21_{-0.21}^{+0.24}$ } & $-16.44_{-1.58}^{+0.03}$  & $9.0\times10^{-3}-1.6\times10^{-1}$\\
Nugent Ibc bright & 2 & $0.50_{-0.03}^{+0.03}$  & $+0.0_{-0.5}^{+0.5}$  & \multicolumn{2}{c}{$0.22_{-0.22}^{+0.02}$ } & $-18.45_{-0.03}^{+0.03}$  & $1.0\times10^{-2}-1.2\times10^{-2}$\\
\hline \hline \multicolumn{8}{c}{Cand-25 ($34.11$~arcmin, photo-$z=0.33^{+0.18}_{-0.19}$)} \\ \hline
Nugent IIP$^{\rm e}$ & 16 & $0.45_{-0.03}^{+0.08}$  & $-0.5_{-1.0}^{+1.5}$  & \multicolumn{2}{c}{$0.00_{-0.00}^{+0.01}$ } & $-17.36_{-0.53}^{+0.08}$  & $1.5\times10^{-5}-6.4\times10^{-4}$\\
\hline \hline \multicolumn{8}{c}{Cand-26 ($33.91$~arcmin, photo-$z=0.31^{+0.03}_{-0.02}$)} \\ \hline
Nugent Ibc normal$^{\rm e}$ & 3 & $0.30_{-0.03}^{+0.03}$  & $-4.8_{-0.5}^{+0.5}$  & \multicolumn{2}{c}{$0.00_{-0.00}^{+0.01}$ } & $-16.60_{-0.08}^{+0.08}$  & $1.4\times10^{-6}-2.9\times10^{-6}$\\
\hline \hline \multicolumn{8}{c}{Cand-27 ($48.41$~arcmin, photo-$z=0.64^{+0.08}_{-0.07}$)} \\ \hline
Nugent Ibc bright$^{\rm e}$ & 73 & $0.65_{-0.08}^{+0.08}$  & $-42.4_{-78.7}^{+6.7}$  & \multicolumn{2}{c}{$0.00_{-0.00}^{+0.01}$ } & $-20.15_{-0.18}^{+0.43}$  & $3.7\times10^{-4}-2.0\times10^{-2}$\\
\hline \hline \multicolumn{8}{c}{Cand-28 ($25.48$~arcmin, photo-$z=0.41^{+0.09}_{-0.07}$)} \\ \hline
Hsiao Ia$^{\rm e}$ & 72 & $0.35_{-0.03}^{+0.03}$  & $+10.8_{-1.5}^{+0.5}$  & $1.10_{-0.31}^{+0.11}$  & $0.00_{-0.13}^{+0.08}$  & $-19.68_{-0.10}^{+0.75}$  & $8.3\times10^{-3}-8.1\times10^{-1}$\\
PS1-10bjp & 4 & $0.35_{-0.03}^{+0.08}$  & $+9.9_{-0.6}^{+0.5}$  & \multicolumn{2}{c}{$0.26_{-0.26}^{+0.02}$ } & $-17.20_{-0.03}^{+0.03}$  & $1.8\times10^{-2}-4.6\times10^{-1}$\\
PS1-10ah & 3 & $0.35_{-0.03}^{+0.03}$  & $+9.9_{-0.5}^{+0.5}$  & \multicolumn{2}{c}{$0.14_{-0.14}^{+0.02}$ } & $-16.63_{-0.03}^{+0.03}$  & $1.9\times10^{-2}-1.5\times10^{-1}$\\
\hline \hline \multicolumn{8}{c}{Cand-29 ($27.95$~arcmin, photo-$z=0.73^{+0.17}_{-0.12}$)} \\ \hline
Nugent Ibc bright$^{\rm e}$ & 20 & $0.65_{-0.03}^{+0.08}$  & $-6.4_{-2.3}^{+4.5}$  & \multicolumn{2}{c}{$0.00_{-0.00}^{+0.01}$ } & $-18.90_{-0.58}^{+0.33}$  & $5.4\times10^{-3}-4.2\times10^{-1}$\\
Hsiao Ia & 307 & $0.90_{-0.28}^{+0.03}$  & $-13.3_{-7.5}^{+10.3}$  & $0.95_{-0.31}^{+0.26}$  & $-0.15_{-0.08}^{+0.63}$  & $-19.74_{-0.39}^{+1.30}$  & $5.2\times10^{-3}-9.4\times10^{-2}$\\
\hline \hline \multicolumn{8}{c}{Cand-30 ($30.75$~arcmin, photo-$z=0.67^{+0.06}_{-0.05}$)} \\ \hline
PS1-10ah$^{\rm e}$ & 4 & $0.65_{-0.03}^{+0.08}$  & $+7.1_{-0.7}^{+0.5}$  & \multicolumn{2}{c}{$0.00_{-0.00}^{+0.01}$ } & $-17.78_{-0.23}^{+0.03}$  & $2.0\times10^{-5}-5.6\times10^{-5}$\\
\hline \hline \multicolumn{8}{c}{Cand-31 ($18.65$~arcmin, photo-$z=0.34^{+0.02}_{-0.03}$)} \\ \hline
Nugent Ibc bright$^{\rm e}$ & 15 & $0.35_{-0.03}^{+0.03}$  & $-81.6_{-2.5}^{+3.5}$  & \multicolumn{2}{c}{$0.00_{-0.00}^{+0.01}$ } & $-20.30_{-0.03}^{+0.18}$  & $1.3\times10^{-3}-1.7\times10^{-2}$\\
Nugent IIn & 241 & $0.35_{-0.03}^{+0.03}$  & $-70.5_{-45.5}^{+5.5}$  & \multicolumn{2}{c}{$0.00_{-0.00}^{+0.01}$ } & $-18.86_{-1.83}^{+0.43}$  & $1.2\times10^{-3}-1.6\times10^{-2}$\\
\hline \hline \multicolumn{8}{c}{Cand-32 ($21.76$~arcmin, photo-$z=0.54^{+0.10}_{-0.06}$)} \\ \hline
Nugent Ibc normal$^{\rm e}$ & 58 & $0.50_{-0.03}^{+0.13}$  & $+8.0_{-3.1}^{+0.7}$  & \multicolumn{2}{c}{$0.00_{-0.00}^{+0.01}$ } & $-17.55_{-0.53}^{+0.58}$  & $4.5\times10^{-3}-1.6\times10^{-1}$\\
Nugent Ibc bright & 151 & $0.50_{-0.03}^{+0.13}$  & $+8.0_{-4.3}^{+1.5}$  & \multicolumn{2}{c}{$0.28_{-0.28}^{+0.28}$ } & $-18.45_{-0.03}^{+0.03}$  & $4.0\times10^{-3}-1.6\times10^{-1}$\\
Nugent IIL normal & 57 & $0.55_{-0.08}^{+0.08}$  & $-1.5_{-1.0}^{+2.5}$  & \multicolumn{2}{c}{$0.54_{-0.54}^{+0.14}$ } & $-17.91_{-0.03}^{+0.03}$  & $4.1\times10^{-3}-1.8\times10^{-2}$\\
PS1-12brf & 47 & $0.50_{-0.03}^{+0.08}$  & $+1.8_{-0.6}^{+2.5}$  & \multicolumn{2}{c}{$0.30_{-0.30}^{+0.05}$ } & $-17.39_{-0.03}^{+0.03}$  & $4.0\times10^{-3}-9.0\times10^{-3}$\\
\hline
\end{longtable}

\normalsize
\begin{longtable}{ccc}
 \caption{Parameters of templates.}\label{tab:template}
 \hline
 Parameter & Range & Step\\
 \hline
 \hline
 \endfirsthead
 \hline
 Parameter & Range & Step\\
 \hline
 \hline
 \endhead
 \endfoot
 \endlastfoot
   \hline \multicolumn{3}{c}{Type Ia SN}\\ \hline
   Color $c$& $-0.2-+0.8$ & $0.05$ \\
   Stretch $s$& $0.6-1.2$ & $0.01$ \\
   Intrinsic variation $I$& $-0.3-+0.3$ & $0.05$ \\
   Redshift $z$& $0-2$ & $0.05$ \\
   \hline \hline
   \multicolumn{3}{c}{Type IIP SN}\\ \hline
   Peak magnitude $M_{\rm B}$& $-14.43--18.91$ & $0.05$ \\
   Color excess $\Ebvh$& $0.0-1.0$ & $0.01$ \\
   Redshift $z$& $0-2$ & $0.05$ \\
   \hline \hline
   \multicolumn{3}{c}{Type IIL normal SN}\\ \hline
   Peak magnitude $M_{\rm B}$& $-16.47--17.99$ & $0.05$ \\
   Color excess $\Ebvh$& $0.0-1.0$ & $0.01$ \\
   Redshift $z$& $0-2$ & $0.05$ \\
   \hline \hline
   \multicolumn{3}{c}{Type IIL bright SN}\\ \hline
   Peak magnitude $M_{\rm B}$& $-17.92--19.96$ & $0.05$ \\
   Color excess $\Ebvh$& $0.0-1.0$ & $0.01$ \\
   Redshift $z$& $0-2$ & $0.05$ \\
   \hline \hline
   \multicolumn{3}{c}{Type IIn SN}\\ \hline
   Peak magnitude $M_{\rm B}$& $-16.98--20.66$ & $0.05$ \\
   Color excess $\Ebvh$& $0.0-1.0$ & $0.01$ \\
   Redshift $z$& $0-4$ & $0.05$ \\
   \hline \hline
   \multicolumn{3}{c}{Type Ibc normal SN}\\ \hline
   Peak magnitude $M_{\rm B}$& $-16.09--17.95$ & $0.05$ \\
   Color excess $\Ebvh$& $0.0-1.0$ & $0.01$ \\
   Redshift $z$& $0-2$ & $0.05$ \\
   \hline \hline
   \multicolumn{3}{c}{Type Ibc bright SN}\\ \hline
   Peak magnitude $M_{\rm B}$& $-18.46--20.30$ & $0.05$ \\
   Color excess $\Ebvh$& $0.0-1.0$ & $0.01$ \\
   Redshift $z$& $0-2$ & $0.05$ \\
   \hline \hline
   \multicolumn{3}{c}{Rapid transient (PS1-10ah)}\\ \hline
   Peak magnitude $M_{\rm B}$& $-16.63--18.63$ & $0.05$ \\
   Color excess $\Ebvh$& $0.0-1.0$ & $0.01$ \\
   Redshift $z$& $0-2$ & $0.05$ \\
   \hline \hline
   \multicolumn{3}{c}{Rapid transient (PS1-10bjp)}\\ \hline
   Peak magnitude $M_{\rm B}$& $-17.20--19.20$ & $0.05$ \\
   Color excess $\Ebvh$& $0.0-1.0$ & $0.01$ \\
   Redshift $z$& $0-2$ & $0.05$ \\
   \hline \hline
   \multicolumn{3}{c}{Rapid transient (PS1-11qr)}\\ \hline
   Peak magnitude $M_{\rm B}$& $-18.03--20.03$ & $0.05$ \\
   Color excess $\Ebvh$& $0.0-1.0$ & $0.01$ \\
   Redshift $z$& $0-2$ & $0.05$ \\
   \hline \hline
   \multicolumn{3}{c}{Rapid transient (PS1-12bv)}\\ \hline
   Peak magnitude $M_{\rm B}$& $-18.44--20.44$ & $0.05$ \\
   Color excess $\Ebvh$& $0.0-1.0$ & $0.01$ \\
   Redshift $z$& $0-2$ & $0.05$ \\
   \hline \hline
   \multicolumn{3}{c}{Rapid transient (PS1-12brf)}\\ \hline
   Peak magnitude $M_{\rm B}$& $-17.39--19.39$ & $0.05$ \\
   Color excess $\Ebvh$& $0.0-1.0$ & $0.01$ \\
   Redshift $z$& $0-2$ & $0.05$ \\
   \hline
\end{longtable}

\end{document}